\documentclass[aps,pre,twocolumn,superscriptaddress,showpacs]{revtex4-1}
\usepackage{amsfonts}
\usepackage{epsfig}
\usepackage{amsmath,amssymb}
\usepackage{times}
\usepackage{xcolor}
\usepackage{multirow}

\begin{document}

\title{Explosive spreading on complex networks: the role of synergy}

\date{\today}

\author{Quan-Hui Liu}
\affiliation{Web Sciences Center, University of Electronic Science and
Technology of China, Chengdu 611731, China}
\affiliation{Big Data Research Center, University of Electronic Science and Technology of China, Chengdu 611731, China}

\author{Wei Wang}
\email{wwzqbx@hotmail.com}
\affiliation{Web Sciences Center, University of Electronic Science and
Technology of China, Chengdu 611731, China}
\affiliation{Big Data Research Center, University of Electronic Science and Technology of China, Chengdu 611731, China}

\author{Ming Tang}
\email{tangminghan007@gmail.com}
\affiliation{Web Sciences Center, University of Electronic Science and
Technology of China, Chengdu 611731, China}
\affiliation{Big Data Research Center, University of Electronic Science and Technology of China, Chengdu 611731, China}

\author{Tao Zhou}
\affiliation{Web Sciences Center, University of Electronic Science and
Technology of China, Chengdu 611731, China}
\affiliation{Big Data Research Center, University of Electronic Science and Technology of China, Chengdu 611731, China}

\author{Ying-Cheng Lai}
\affiliation{School of Electrical, Computer and Energy Engineering, Arizona State University, Tempe, Arizona 85287, USA}

\begin{abstract}

In spite of the vast literature on spreading dynamics on complex networks, the
role of local synergy, i.e., the interaction of elements that when combined
produce a total effect greater than the sum of the individual elements,
has been studied but only for irreversible spreading dynamics. Reversible
spreading dynamics are ubiquitous but their interplay with synergy has
remained unknown. To fill this knowledge gap, we articulate a model to
incorporate local synergistic effect into the classical
susceptible-infected-susceptible process, in which the probability for a
susceptible node to become infected through an infected neighbor is enhanced
when the neighborhood of the latter contains a number of infected nodes. We
derive master equations incorporating the synergistic effect,
with predictions that agree well with the numerical results. A striking
finding is that, when a parameter characterizing the strength of the synergy
reinforcement effect is above a critical value, the steady state density of
the infected nodes versus the basic transmission rate exhibits an explosively
increasing behavior and a hysteresis loop emerges. In fact,
increasing the synergy strength can promote the spreading and reduce the
invasion and persistence thresholds of the hysteresis loop. A physical
understanding of the synergy promoting explosive spreading and the associated
hysteresis behavior can be obtained through a mean-field analysis.

\end{abstract}


\maketitle

\section{Introduction} \label{sec:goal}

Disease or information spreading, a fundamental class of dynamical processes
on complex networks~\cite{BBV:book,Castellano:2009,Newman:book,Pastor-Satorras:2015}, has been studied extensively
in the past fifteen years~\cite{Pastor-Satorras:2001,Newman:2002,Zanette:2002,LLY:2003,BBPSV:2004,Small:2005,
ZLL:2007,YHL:2008,TLL:2009,Gross:2006,KGHLMSM:2010,YWLXW:2011,ZFC:2012,BH:2013,Glesson:2013,SGR:2014,GGA2013,WTYDLL:2014,Liu:2016}. Spreading dynamics can be classified into two types: irreversible and reversible. In an irreversible
process, once an individual becomes infected, it cannot recover or return to
the susceptible state. Or, once an infected node recovers, it is immune to
the same virus. Mathematically, irreversible spreading processes can be
described by the susceptible-infected (SI), the susceptible-infected-recovered
(SIR)~\cite{Newman:2002}, or the susceptible-exposed-infected-recovered
(SEIR) model~\cite{Small:2005}. In contrast, in a reversible process,
any node can be infected repeatedly in time, going through a cycle of
susceptible and infected states. For example, in the infection process of
tuberculosis and gonorrhea, an individual recovering from such a disease
can be infected again with the same disease anytime. Mathematically,
reversible spreading processes can be described by the
susceptible-infected-susceptible (SIS)~\cite{Pastor-Satorras:2001}, the
susceptible-infected-recovered-susceptible (SIRS)~\cite{Bancal:2010}, or
the susceptible-exposed-recovered-susceptible (SEIS) model~\cite{Masuda:2006}.
When the complex topology of the underlying network is taken into account,
a pioneering result was the vanishing epidemic threshold in scale-free
networks with the power-law exponent less than
three~\cite{Pastor-Satorras:2001}. Another result is that, for both
irreversible and reversible processes described by the classic SIR and SIS
models, respectively, the fraction of infected nodes increases with the
transmission rate continuously~\cite{Pastor-Satorras:2015}, which can
be expected intuitively.

In this paper, we investigate the effect of synergy on reversible spreading
dynamics on complex networks. Synergy describes the situation where the
interaction of elements that produce a total effect greater than the sum
of individual elements when combined, i.e., the phenomenon commonly known as
``one plus one is greater than two.'' Intuitively, synergy should have a
significant effect on spreading dynamics. For example, when a disease
begins to spread in the human society, a healthy individual who has a sick
friend is likely to be infected with the disease. However, if the sick friend
himself or herself has a number of friends with the same disease, the
likelihood for the healthy individual to contract the disease would be higher,
as (a) the fact that his/her sick friend has sick friends implies that the
disease is potentially more contagious, and (b) the healthy individual is
likely to have more sick friends. Similarly, in rumor or information spreading
over a social network, a number of connected individuals possessing a piece
of information make it more believable than just a single individual.
Indeed, concrete evidence existed in both biological and social systems
where the number of infected neighbors of a pair of infected-susceptible
nodes would enhance the transmission rate between them~\cite{Granovetter:1978,
Watts:2002,Lockwood:1988,Jonathan:2011}, such as fungal infection in
soil-borne plant pathogens~\cite{Lockwood:1988,Jonathan:2011} where the
probability for an infected node to affect its susceptible neighbors depends
upon the number of other infected nodes connected to the infected node. In
social systems, the synergistic effect was deemed important in phenomena such
as the spread of adoption of healthy behavior~\cite{Centola:2010,Wang:2015},
microblogging retweeting~\cite{Hodas:2014}, opinion spreading and
propagation~\cite{Castellano:2009,LCZ:2011}, and animal invasion~\cite{Murray:2002,
Gordon:2010}.

While the classic SIR and SIS models ignore the synergistic effect by assuming
that the transmission of infection between a pair of infected-susceptible
nodes is independent of the states of their neighbors, there were previous
efforts to study the impact of synergy on {\em irreversible} spreading
dynamics and its interplay with the network topology. In particular, threshold
models~\cite{Granovetter:1978,Watts:2002,GLM:2001} were developed, which take into
account neighbors' synergistic effects on behavior spreading by assuming that
a node adopts a behavior only when the number of its adopted neighbors is
equal to or exceeds a certain adoption threshold. One result was that, for
each node in the network with a fixed adoption threshold, the final adoption
size tends to grow continuously and then decreases discontinuously when the mean degree of the network is increased. The SIR model was also
generalized to modify the transmission rate between a pair of infected and
susceptible nodes according to the synergistic effect~\cite{Perez-Reche:2011,
Taraskin:2013,Broder-Rodgers:2015}, with the finding that it can affect
the fraction of the epidemic outbreak, duration and foraging strategy of
spreaders. These existing works were exclusively for irreversible
spreading dynamics. A systematic study to understand the impact of the
synergistic effects on {\em reversible} spreading dynamics on complex
networks is needed.

The goal of this paper is to investigate, analytically and numerically, the
impacts of synergy on reversible spreading dynamics on complex networks. We
first generalize the classic SIS model to quantify the effect of the number of
infected neighbors connected to an infected node on the transmission rate
between it and its susceptible neighbors. To characterize the impact on
the steady state of the spreading dynamics, we consider the local nodal
environment and derive the master equations (MEs)~\cite{Lindquist:2011,
Glesson:2011}. To gain a physical understanding, we assume that,
statistically, nodes with the same degree have the same dynamical
characteristics, so the mean-field approximation can be applied. Let $\alpha$
be a parameter characterizing the strength of the synergistic effect.
For random regular networks (RRNs), we find that for $\alpha \geq \alpha_c$, where
$\alpha_c$ is a critical value, a hysteresis loop~\cite{Gross:2006,Yang:2015}
appears in which the steady state infected density, denoted by $\rho(\infty)$,
increases with the transmission rate $\beta$ but typically exhibits an
explosively increasing behavior, in contrast to the
typical continuous transition observed in the classic SIS
models~\cite{Pastor-Satorras:2001}. For $\alpha < \alpha_c$, the hysteresis
loop disappears and $\rho(\infty)$ increases with $\beta$ continuously.
The phenomena of explosive spreading and hysteresis loop are general in that they also occur for complex networks of different topologies.

\section{Model} \label{sec:model}

\paragraph*{Network model.}
The networks in our study are generated from the uncorrelated configuration
model~\cite{Newman:2002} with degree distribution $P(k)$, where the
degree-degree correlations can be neglected for large and sparse networks.
Nodes in the network correspond to individuals or hosts responsible for
spreading, with edges representing the interactions between nodal pairs.

\begin{figure}[!htb]
\centering
\includegraphics[width=\linewidth]{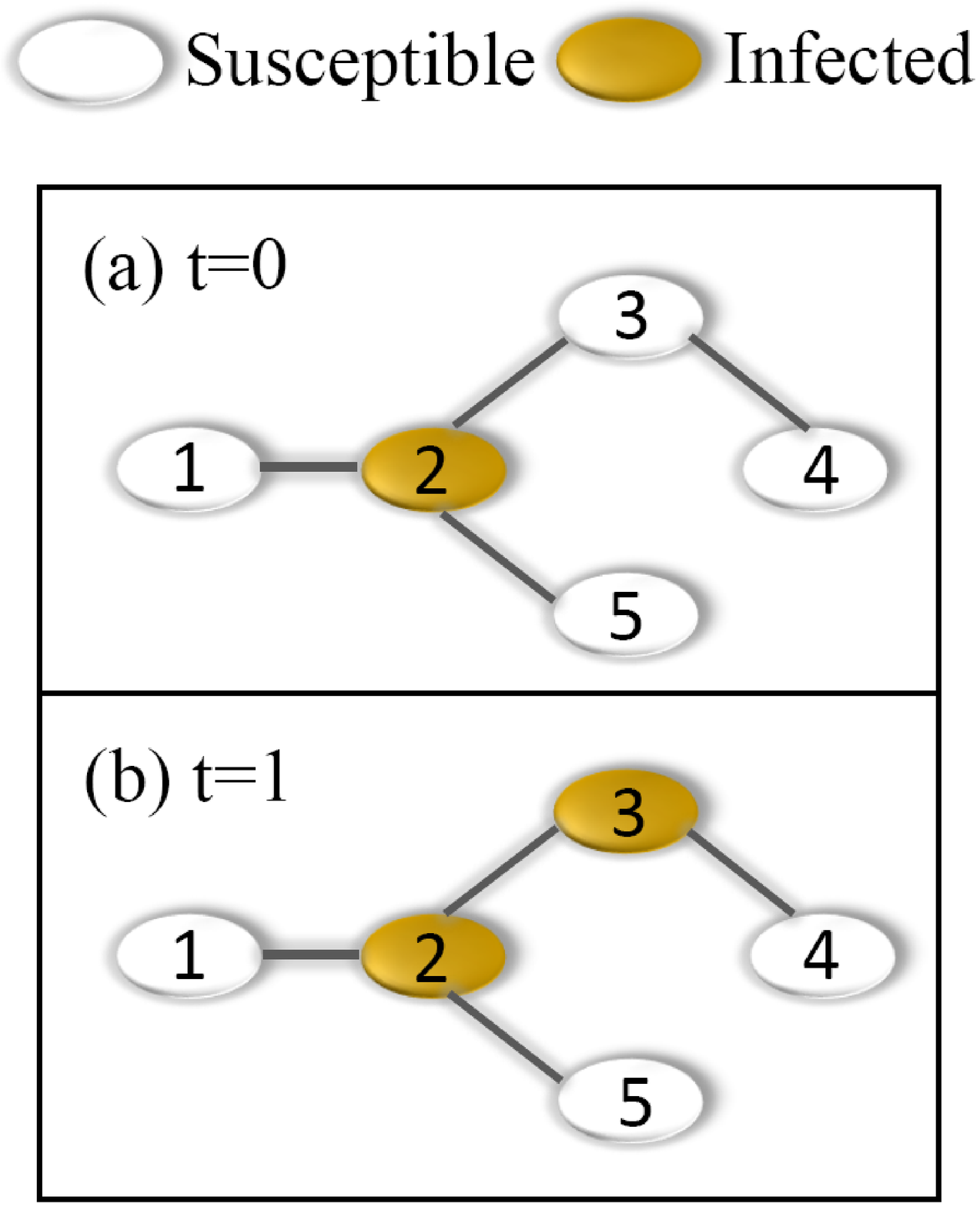}
\caption{ (Color online) Illustration of synergistic SIS
spreading process on complex networks. (a) Initially (at $t=0$), node 2 is
the seed and the remaining nodes are susceptible. Since there are
no infected neighbors connected to node 2, it transmits the disease to one
of its susceptible neighbors with probability $p(0,\alpha)=\beta$.
(b) Node 3 is infected by node 2 which has not recovered. In this case,
both nodes 2 and 3 have an infected neighbor and, at the next time step,
they will infect one of their susceptible neighbors with a larger probability
$p(1,\alpha)\simeq(1+\alpha)\beta$ due to the synergistic effect.}
\label{fig:schematic}
\end{figure}

\paragraph*{Model of reversible spreading dynamics.}
We generalize the classic SIS model to incorporate the synergistic effect
into the reversible spreading dynamics --- we name it the {\em synergistic
SIS spreading model}. At any time, each node can only be in one of two states:
susceptible (S) or infected (I). An infected node can transmit the disease
to its susceptible neighbors. The synergistic mechanism models the role of
infected neighbors connected to a transmitter (i.e., an infected node) in
enhancing the transmission probability. The synergistic SIS spreading
process is illustrated schematically in Fig.~\ref{fig:schematic}. Our
model differs from the recent one in Ref.~\cite{Gomez:2016}, which treated
the synergistic effect of ignorant individuals attached to a receiver (in ignorant state).

Initially, a fraction $\rho_0$ of nodes are chosen as seeds (infected nodes)
at random, while the remaining nodes are in the susceptible state. Each
infected node can transmit the disease to its susceptible neighbors at
the rate
\begin{eqnarray} \label{eq:PM}
p(m,\alpha)=1-(1-\beta)^{1+{\alpha}m},
\end{eqnarray}
where $m$ and $\alpha$, respectively, represent the number of the infected
neighbors connected to the infected node and the strength of the synergistic
effect, and $\beta$ is the basic transmission rate. Equation~\eqref{eq:PM}
indicates that, the larger value of $\alpha$ or \emph{m}, the higher the
transmission rate $p(m,\alpha)$ between an infected node and a susceptible
neighbor will be. An infected node can recover to being susceptible with
probability $\mu$. Our model reduces to the classic SIS model for $\alpha=0$.
For $\alpha>0$ ($\alpha<0$), the synergistic effects are constructive
(destructive) where the infected neighbors favor (hampers) transmission of
the disease to the receivers. In our study, we consider only the
constructive synergistic effect, where the infected neighbors of an infected
node cooperate with it to spread the disease. In addition, we set
$\alpha \leq 1$ so that the synergistic ability of any infected neighbor of
the infected node is less than that of itself. This assumption is based on
consideration of real situations such as fungal infection in soil-borne
plant pathogens where the probability for a susceptible node infected by
a direct infected neighbor is always greater than that from an indirect
infected neighbor~\cite{Lockwood:1988,Jonathan:2011}.

\section{Theory} \label{sec:theory}

We consider large and sparse networks with negligible degree-degree
correlation. We first establish the master equations to describe the
synergistic SIS spreading process quantitatively. We then provide an
an intuitive understanding of the role of synergy in the spreading
dynamics through a mean-filed analysis.

\subsection{Master equations} \label{subsec:ME}

In general, the transmission rate $p(m,\alpha)$ between a pair of
infected-susceptible nodes in the synergistic SIS spreading process
is determined by the following three factors: (1) the basic transmission
rate $\beta$ between the pair of nodes, i.e., the rate in the absence
of any synergistic effect, (2) the number of infected neighbors
connected to the infected node, and (3) the strength $\alpha$ of the
synergistic effect. Because of the strong dynamical correlation among
the states of the neighboring nodes leading to the synergistic effect,
the approach of master equations~\cite{Lindquist:2011,Glesson:2011} can
be applied. For convenience, we denote $S_{k,m}$ ($I_{k,m}$) as the
k-degree susceptible (infected) node with $m$ infected neighbors and
use $s_{k,m}(t)$ and $i_{k,m}(t)$ to express the fractions of $S_{k,m}$
and $I_{k,m}$ nodes at time $t$, respectively. The degree distribution
and the average degree of the network are $P_{k}$ and
${\langle k \rangle}=\sum_{k^\prime}k^{\prime}P_{k^{\prime}}$,
respectively. The fraction of infected nodes with degree $k$ at time $t$
is given by
\begin{displaymath}
\rho_{k}(t)=\sum_{m=0}^{k}i_{k,m}(t)=1-\sum_{m=0}^{k}s_{k,m}(t),
\end{displaymath}
and the total fraction of the infected nodes is
$\rho(t)={\langle\rho_{k}(t)\rangle}\equiv{\sum_{k}P_{k}\rho_{k}(t)}$.

To derive the master equations, it is necessary to obtain the probability
for $S_{k,m}$ to be infected. Initially, $S_{k,m}$ has $m$ infected
neighbors so the probability for one of its infected neighbors to have degree
$k^{\prime}$ is $k^{\prime}P_{k^{\prime}}/\langle k \rangle$. This
degree $k^{\prime}$ infected neighbor of $S_{k,m}$ may have zero, one, two,
or up to $k^{\prime}-1$ infected neighbors. The chance for the
degree $k^{\prime}$ infected node to have $n$ infected neighbors is
$i_{k^{\prime},n}(t)/i_{k^{\prime}}(t)$, so the probability that it will
infect $S_{k,m}$ is
\begin{displaymath}
\sum_{n=0}^{k^{\prime}-1}\frac{i_{k^{\prime},n}(t)}{i_{k^{\prime}}(t)}
p(n,\alpha).
\end{displaymath}
Since $S_{k,m}$ has $m$ infected neighbors, the probability of
its being infected during time $t+dt$, where $dt$ is an infinitesimally
small time interval, can be written as $\pi_{k,m}(t)dt$ with $\pi_{k,m}(t)$
given by
\begin{eqnarray} \label{eq:Pi}
\pi_{k,m}(t)=m\sum_{k^{\prime}}\frac{k^{\prime}P_{k^{\prime}}}
{\langle k \rangle}\sum_{n=0}^{k^{\prime}-1}
\frac{i_{k^{\prime},n}(t)}{i_{k^{\prime}}(t)}p(n,\alpha).
\end{eqnarray}
There are three scenarios that can lead to an increase in $s_{k,m}(t)$:
(1) recovery of $I_{k,m}$ with probability $\mu$, (2) infection
of a susceptible neighbor of $S_{k,m-1}$, and (3) recovery of an
infected neighbor of $S_{k,m+1}$. The second (third) scenario
corresponds to the situation where an S-S (S-I) edge changes
into an S-I (S-S) edge, where an S-S edge connects
two susceptible nodes, an S-I edge links a susceptible and an
infected nodes, and so on. Denote $\beta^s$ as the rate that an S-S
edge changes to S-I. We can approximate $\beta^s$ as the rate of edges
that switch from being S-S to S-I in the time interval $dt$, and the
probability $\beta^sdt$ is the ratio of the latter to the former. The
rate $\beta^s$ can thus be approximated as
\begin{eqnarray} \label{eq:BetaS}
\beta^s=\frac{\sum{P_k}\sum_{m=0}^{k}(k-m)\pi_{k,m}(t)s_{k,m}(t)}
{\sum{P_k} \sum_{m=0}^{k}(k-m)s_{k,m}(t)}.
\end{eqnarray}
Since the probability for the recovery of an infected node does not depend
on its neighbors, the rate at which an S-I edge changes to S-S is $\mu$.
Similarly, there are three cases leading to a decrease in $s_{k,m}(t)$:
$S_{k,m}$ being infected with probability $\pi_{k,m}$, infection of a
susceptible neighbor of $S_{k,m}$ with probability $\beta^{s}$, and
recovery of an infected neighbor of $S_{k,m}$ with probability ${\mu}$.
We then obtain the time evolution equation of $s_{k,m}(t)$ as
\begin{eqnarray}\label{eq:DiffSkm}
\frac{d}{dt}s_{k,m}(t)&=&
\nonumber
{\mu}i_{k,m}(t)+{\beta^{s}}(k-m+1)s_{k,m-1}(t)\\
\nonumber
&+&\mu(m+1)s_{k,m+1}(t)\\
&-&[\pi_{k,m}(t)+\beta^{s}(k-m)+{\mu}m]s_{k,m}(t),
\end{eqnarray}
Analogously, we can derive the time evolution equation of $i_{k,m}(t)$:
\begin{eqnarray} \label{eq:DiffIkm}
\frac{d}{dt}i_{k,m}(t)&=&
\nonumber
\pi_{k,m}(t)s_{k,m}(t)+{\beta^{i}}(k-m+1)i_{k,m-1}(t)\\
\nonumber
&+&\mu(m+1)i_{k,m+1}(t)\\
&-&[\mu+\beta^{i}(k-m)+{\mu}m]i_{k,m}(t),
\end{eqnarray}
where $\beta^i$ is the rate with which an edge S-I switches to I-I,
which can be calculated as
\begin{eqnarray} \label{eq:BetaI}
\beta^i=\frac{\sum{P_k} \sum_{m=0}^{k}m\pi_{k,m}(t)s_{k,m}(t)}
{\sum{P_k} \sum_{m=0}^{k}m{s_{k,m}(t)}}.
\end{eqnarray}
If the initially infected nodes are distributed uniformly on the network,
the initial conditions of Eqs.~(\ref{eq:Pi})-(\ref{eq:BetaI}) are
\begin{eqnarray}
\nonumber
s_{k,m}(0) & = & [1-\rho(0)]B_{k,m}[\rho(0)]~and \\ \nonumber
i_{k,m}(0) & = & \rho(0)B_{k,m}[\rho(0)],
\end{eqnarray}
where $B_{k,m}(p)=\binom{k}{m}{p}^m{(1-p)}^{k-m}$. Numerically solving
Eqs.~(\ref{eq:Pi})-(\ref{eq:BetaI}), we obtain the quantities $i_{k,m}$
and $s_{k,m}$ at any time $t$. The quantity $\rho(\infty)$ can be
calculated as $\rho(\infty)=\sum_k{P_k}\sum_{m=0}^{m=k}i_{k,m}(\infty)$,
and we have $s(\infty)=1-\rho(\infty)$. For simplicity, we denote $\rho(\infty)=\rho$.

\subsection{Mean-field approximation} \label{subsec:MFT}

To gain physical insights into the role of synergistic effects in spreading
dynamics, we develop a mean-field analysis. In particular, we assume
that nodes with the same degree exhibit approximately identical dynamical
behaviors. The time evolution of the fraction of the degree $k$ infected
nodes is then given by
\begin{eqnarray}\label{eq:DiffRhoK}
\frac{d}{dt}\rho_k(t)
\nonumber
&=&[1-\rho_k(t)]k  \\
\nonumber
&\times&\sum_{k^{\prime}}\frac{k^{\prime}P_{k^{\prime}}
\rho_{k^{\prime}}}{\langle k \rangle}\sum_{m=0}^{k^{\prime}-1}
B_{{k^{\prime}-1},m}(w)p(m,\alpha)\\
&-&\mu\rho_k(t),
\end{eqnarray}
where $w=\sum{k{P_k}{\rho_k}}/{\langle k \rangle}$ is the probability
that one end of a randomly chosen edge is infected,
$\rho(t)={\sum}P_{k}\rho_{k}(t)$, and the fraction of susceptible nodes
at time $t$ is $s(t)=1-\rho(t)$. The steady state of synergistic SIS
process in Eq.~(\ref{eq:DiffRhoK}) corresponds to the condition
$\frac{d}{dt}\rho_k(t)=0$. For degree $k$ we have
\begin{eqnarray} \label{eq:RhoEq}
\rho_k(\infty) & = &
\nonumber
\frac{[1-\rho_k(\infty)]k}{\mu}\\
&\times& \sum_{k^{\prime}}\frac{k^{\prime}P_{k^{\prime}}
\rho_{k^{\prime}}(\infty)}{\langle k \rangle}\sum_{m=0}^{k^{\prime}-1}
B_{{k^{\prime}-1},m}(w)p(m,\alpha),
\end{eqnarray}
which can be solved analytically for RRNs by approximating
$1-{(1-\beta)}^{(1+{\alpha}m)}$ as $\beta(1+{\alpha}m)$ for small
$\beta$. We get
\begin{eqnarray} \label{eq:RhoInfety}
\nonumber
\rho(\infty) &=& -\frac{{\alpha}{\beta}k(k-1)}{\mu}{\rho(\infty)}^3
+\frac{[\alpha{\beta}k(k-1)
-{\beta}k]}{\mu}{\rho(\infty)}^2\\
&+&\frac{{\beta}k}{\mu}{\rho(\infty)},
\end{eqnarray}
for $t\rightarrow\infty$. Solving Eq.~(\ref{eq:RhoInfety}), we get
the infected density $\rho(\infty)$.

\begin{figure}[!htb]
\centering
\includegraphics[width=\linewidth]{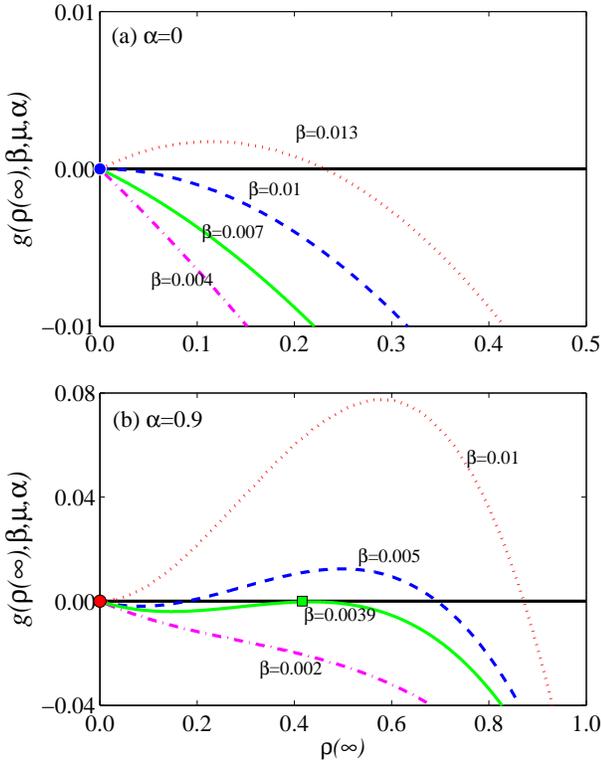}
\caption{(Color online) Illustration of graphical solution of
Eq.~\eqref{eq:GFunction}. For random regular networks with $k=10$,
(a) continuously increasing behavior of $\rho(\infty)$ with $\beta$
for $\alpha=0$, (b) explosive change in $\rho(\infty)$ for $\alpha=0.9$.
The blue dashed line is tangent to the horizontal axis at $\rho(\infty)=0$
(i.e., the blue circle) in (a). The red circle and green square
respectively represent the points of tangency for the red dotted line
and green solid line in (b). The recovery rate is $\mu=0.1$.}
\label{fig:RRN_GF}
\end{figure}

The epidemic threshold is a critical parameter value above which a
global epidemic occurs but below which there is no epidemic. Similar to
the analysis of the classic SIS spreading dynamics, we can obtain the
critical condition from the nontrivial solution of Eq.~(\ref{eq:RhoInfety}).
In particular, the function
\begin{eqnarray} \label{eq:GFunction}
g[\rho(\infty),\beta,\mu,\alpha]&=&
\nonumber -\frac{{\alpha}{\beta}k(k-1)}{u}{\rho(\infty)}^3 \\
\nonumber &+&\frac{[\alpha{\beta}k(k-1)-{\beta}k]}{\mu}{\rho(\infty)}^2 \\
&+&\frac{{\beta}k}{\mu}{\rho(\infty)}-{\rho(\infty)},
\end{eqnarray}
becomes tangent to the horizontal axis at $\rho_c(\infty)$, which is the critical
infected density in the limit $t\to\infty$. The critical condition is given by
\begin{eqnarray} \label{eq:DiffRho}
\frac{dg[\rho(\infty),\beta,\mu,\alpha]}{d\rho_{\infty}}\rvert_{\rho_c(\infty)}=0.
\end{eqnarray}
Furthermore, the basic critical transmission rate can be calculated as:
\begin{eqnarray} \label{eq:BetaC}
\beta_c=\frac{\mu}{\Gamma},
\end{eqnarray}
where
\begin{displaymath}
\Gamma=k[1-2(1-(k-1)\alpha)\rho_c(\infty)-3(k-1)\alpha{\rho_c(\infty)}^{2}].
\end{displaymath}
Numerically solving Eqs.~(\ref{eq:RhoInfety}) and (\ref{eq:BetaC}), we
get the critical transmission rate $\beta_c$. For $\alpha=0$, our
synergistic SIS spreading model reduces to the classic SIS spreading
model, and Eq.~(\ref{eq:RhoInfety}) has a trivial solution $\rho(\infty)=0$.
For $\alpha=0$, Eq.~(\ref{eq:RhoInfety}) has only one nontrivial solution.
We thus see that $\rho(\infty)$ increases with $\beta$ continuously. As
shown in Fig.~\ref{fig:RRN_GF}(a), the function
$g[\rho(\infty),\beta,\mu,\alpha]$ is tangent to the horizontal axis at
$\rho(\infty)=0$. Combining Eqs.~(\ref{eq:RhoInfety}) and (\ref{eq:BetaC}),
we obtain the continuous critical transmission rate $\beta_c=\mu/k$ for
$\alpha=0$.

For $\alpha > 0$ so synergistic effects exist, $\rho({\infty})=0$ is
a trivial solution since Eq.~(\ref{eq:RhoInfety}) is a cubic equation for
the variable $\rho(\infty)$ without any constant term. As shown in
Fig.~\ref{fig:RRN_GF}(b), for a fixed $\alpha> 0$ (e.g., $\alpha=0.9$), the
number of solutions of Eq.~(\ref{eq:RhoInfety}) is dependent upon
$\beta$, and there exists a critical value of $\beta$ at which
Eq.~(\ref{eq:RhoInfety}) has three roots (fixed points), indicating
the occurrence of a saddle-node bifurcation~\cite{Ott:book,Strogatz:1994}. The bifurcation analysis of Eq.~(9) reveals the physically meaningful stable solution of $\theta(\infty)$ will suddenly increase to an alternate outcome. In this case, an explosive growth pattern of $\rho(\infty)$ with $\beta$ emerges. And whether the unstable state stabilizes to an outbreak
state [$\rho(\infty)>0$] or an extinct state [$\rho(\infty)=0$] depends
on the initial fraction of the infected seeds. As a result, a hysteresis
loop emerges~\cite{Gross:2006,Yang:2015}. To distinguish
the two thresholds of the hysteresis loop, we denote $\beta_{inv}$ as
the invasion threshold corresponding to the trivial solution
[$\rho(\infty)=0$] of Eq.~(\ref{eq:RhoInfety}), associated with which the
disease starts with a small initial fraction of the infected seeds, and
let $\beta_{per}$ be the persistence threshold corresponding to the
nontrivial solution [$\rho_c(\infty)>0$] of Eq.~(\ref{eq:RhoInfety}), at
which the disease starts with a higher initial fraction of the infected
seeds~\cite{Gross:2006,Yang:2015}. Substituting the trivial solution
[$\rho(\infty)=0$] into Eq.~\eqref{eq:BetaC}, we obtain the invasion
threshold as
\begin{eqnarray} \label{eq:BetaC2}
\beta_{inv}=\frac{\mu}{k}.
\end{eqnarray}
Note that the classic SIS spreading process has the same invasion
threshold. We can also solve Eqs.~(\ref{eq:RhoInfety})
and (\ref{eq:BetaC}) simultaneously to get the persistence threshold
$\beta_{per}$ with $\rho_c(\infty)>0$.

We now present an explicit example to understand the relationship between
$\rho(\infty)$ and $\beta$. As shown in Fig.~\ref{fig:RRN_GF}(b) for
$\alpha=0.9$, numerically solving Eqs.~(\ref{eq:RhoInfety}) and
(\ref{eq:BetaC}) gives the function $g[\rho(\infty),\beta,\gamma,\alpha]$,
which becomes tangent to the horizontal axis for $\beta_{inv}=0.01$ or
$\beta_{per}\approx0.0039$. From Fig.~\ref{fig:RRN_GF}(b),
we see that Eq.~(\ref{eq:RhoInfety}) has $3$ fixed points when $\beta$
is in the range of ($\beta_{inv}, \beta_{per})$. As a result, the steady
state infection density depends on $\rho_0$. If the disease starts with a
small initial fraction of infected seeds, the root with the smallest value
[$\rho(\infty)=0$] of Eq.~(\ref{eq:RhoInfety}) corresponds to the
steady state. However, if the disease starts with a large initial fraction
of infected seeds, the root with the largest value is the infected density
in the steady state. When $\beta$ is smaller than $\beta_{per}$ or larger
than $\beta_{inv}$, the initial fraction of infected seeds has no effect
on the steady state.

Next, by solving the condition of the saddle-node bifurcation~\cite{Ott:book,Strogatz:1994},
we can determine the critical value of infected neighbors' synergy effects $\alpha_c$, for
$\alpha<\alpha_c$, $\rho(\infty)$ increases with $\beta$ continuous, while $\rho(\infty)$ will increase with $\beta$ explosively and the hysteresis appears when $\alpha>\alpha_c$.
Combing Eqs.~(\ref{eq:RhoInfety}) and (\ref{eq:DiffRho}) together with the
condition
\begin{eqnarray} \label{eq:Diff2G}
\frac{d^2g[\rho(\infty),\beta,\mu,\alpha]}{d\rho_{\infty}^2}\rvert_{\rho_c(\infty)}=0,
\end{eqnarray}
we obtain
\begin{eqnarray}\label{eq:AlphaC}
\alpha_c=\frac{1}{k-1-3(k-1)\rho_c{(\infty)}}.
\end{eqnarray}
Combining Eqs.~(\ref{eq:RhoInfety}),~(\ref{eq:DiffRho}) and~(\ref{eq:AlphaC}),
we get $\alpha_c=1/(k-1)$, which is dependent only on the degree of the
RRNs.

\section{Numerical verification}

We perform extensive simulations of synergistic SIS spreading processes
on RRNs of size $N=10^4$ and degree $k=10$. To
calculate the pertinent statistical averages we use 30 network realizations
and at least $10^3$ independent dynamical realizations for each parameter
setting. To be concrete, we take synchronous updating processes~\cite{Pastor-Satorras:2015} and set the recovery rate as $\mu=0.1$ in all
simulations (unless otherwise specified). To obtain the numerical thresholds
$\beta_{inv}$ and $\beta_{per}$, we adopt the susceptibility
measure~\cite{Ferreira:2012,Shu:2015}:
\begin{eqnarray} \label{eq:Chi}
\chi = N\frac{\langle \rho(\infty)^2 \rangle -
{\langle \rho(\infty) \rangle}^2}{\langle \rho(\infty) \rangle},
\end{eqnarray}
where $\rho(\infty)$ is the steady-state density of infected nodes.
In general, $\chi$ exhibits a maximum value at $\beta_{inv}$ and
$\beta_{per}$ when the initial fraction of the infected seeds is
relatively small and large, respectively. We define $\beta_{inv}^{s}$ ($\beta_{per}^{s}$) as the numerical predictions of
invasive (persist) threshold.

\begin{figure}
\begin{center}
\includegraphics[width=\linewidth]{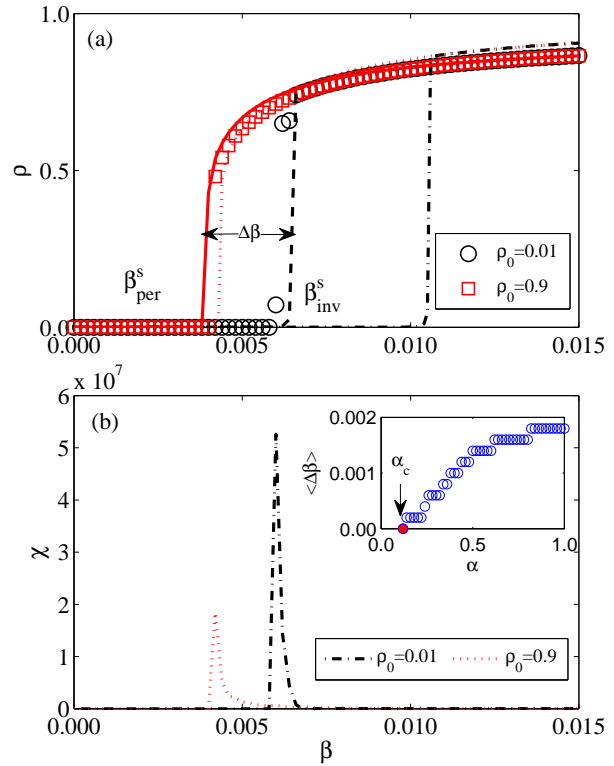}
\caption{(Color online) Steady state infected density $\rho(\infty)$
and susceptibility measure $\chi$ for random regular networks.
(a) The density $\rho(\infty)$ versus $\beta$
for $\alpha=0.9$, where the red squares and black circles are simulation
results with initial infected density $\rho_0=0.9$ and $\rho_0=0.01$,
respectively. The red solid and black dashed lines are the results of
master equations Eqs.~(\ref{eq:BetaS})-(\ref{eq:BetaI}) with the same
respective initial seed fractions. The red dotted and black dotted dashed
lines are results from the mean-field approximation [Eq.~(\ref{eq:DiffRho})]
with the same respective initial seed fractions. The quantities $\beta_{inv}^s$
and $\beta_{per}^s$ are, respectively, the simulated invasion and
persistence thresholds determined via the susceptibility measure.
(b) Susceptibility measure $\chi$ versus $\beta$ with the same parameters
as in (a). To discern the extremely small value of $\chi$ for $\rho_0=0.9$,
we plot the dotted line in (b) one thousand times larger than the original
values. The inset in (b) shows the width of the hysteresis loop versus
$\alpha$. Other parameters are $\mu=0.1$ and $k=10$.}
\label{fig:chi}
\end{center}
\end{figure}

Figure~\ref{fig:chi}(a) shows $\rho(\infty)$ versus $\beta$ for $\alpha=0.9$,
where the surprising phenomenon of explosive spreading, i.e., $\rho(\infty)$
exhibits an explosive increase as $\beta$ passes through a critical point, can
be seen, as predicted [Eqs.~(\ref{eq:Pi})-(\ref{eq:BetaI}), and
Eq.~(\ref{eq:RhoInfety})]. In fact, there exists a range in $\beta$:
[$\beta_{inv}$, $\beta_{per}$], in which the steady state depends on
the value of $\rho_0$. In particular, the two different steady states
correspond to the spreader-free state [$\rho(\infty)=0$] for initially small fraction of infected seeds and the endemic state [$\rho(\infty)>0$] with initially larger fraction of infected nodes, respectively. The coexistence of
endemic and spreader-free states, in the form of a hysteresis loop with
explosive transitions between the states, is predicted by both theoretical
approaches (i.e., the master equations and the mean-field theory), and is
observed numerically. Figure~\ref{fig:chi}(b) shows the susceptibility
measure $\chi$ versus $\beta$ for the two cases of $\rho_0=0.01$ and
$\rho_0=0.9$. We see that the numerical thresholds $\beta_{inv}^s$ and
$\beta_{per}^s$ determined through $\chi$ match well with the predictions
from the master equations, but the mean-field approximation gives only
the value of $\beta_{per}^s$ correctly. Letting $\triangle\beta$ be the
difference between $\beta_{inv}^s$ and $\beta_{per}^s$ (the width of the
hysteresis loop), we find that $\triangle\beta$ increases with $\alpha$,
as shown in the inset of Fig.~\ref{fig:chi}(b), indicating that $\beta_{inv}^s$
decreases faster than $\beta_{per}^s$ as $\alpha$ is increased.

\begin{figure}
\centering
\includegraphics[width=\linewidth]{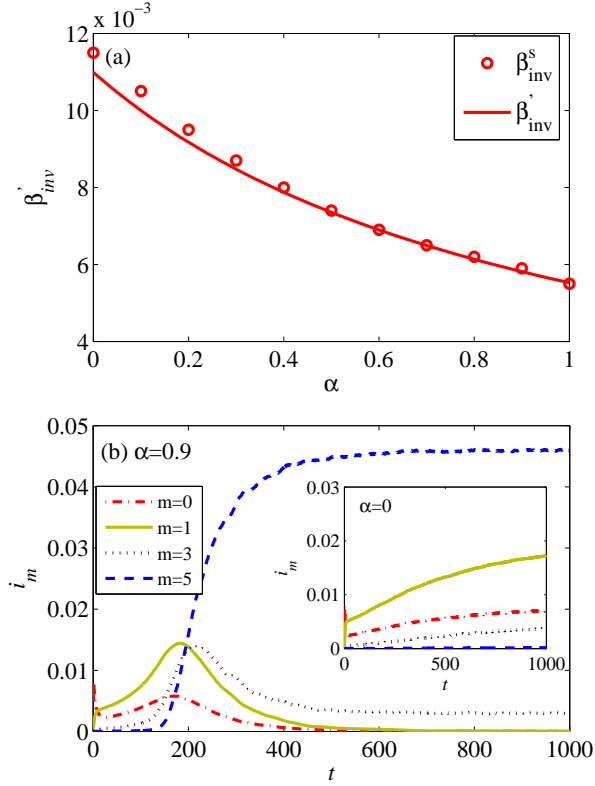}
\caption{(Color online) Illustration the regime of explosive
spreading. (a) Circles indicate the numerical predictions of invasive threshold $\beta^{'}_{inv}$ in $\alpha$. The
solid line shows the transmission rate $\beta$ in Eq.~\eqref{qBeta}. (b) The
fraction $i_m$ of infected nodes for different numbers of infected neighbors
($m=0,1,3,5$) versus time $t$ when the transmission rate is slightly larger
than $\beta_{inv}^s$. Panel (b) shows $i_m$ versus $t$ for
$\alpha=0.9$ and $\beta=0.0064$ ($\beta_{inv}^s=0.0062$), where the inset
shows the same plot for the classic SIS spreading dynamics for
$\beta=0.0114$ ($\beta_c=0.0112$). Other parameters are $\rho_0=0.01$,
$\mu=0.1$ and $k=10$.}
\label{fig:ES_regime}
\end{figure}

To explain why mean-field approximation can't accurately predict $\beta_{inv}^s$,
and to give a qualitative explanation for the explosively increasing behavior of
$\rho(\infty)$ with $\beta$, we consider the case where the spreading
process starts from a small fraction of infected seeds. Initially, for an
infected seed [e.g., node $2$ in Fig.~\ref{fig:schematic}(a)], all
its neighbors are in the susceptible state. Thus, there is no synergistic
effect when this infected node attempts to infect its susceptible neighbors.
Once the infected node ($I_{k,0}$) has infected one of its susceptible
neighbors [e.g., node $3$ in Fig.~\ref{fig:schematic}(a)] successfully, both
the originally and newly infected nodes become $I_{k,1}$, leading to a
synergistic effect. In this case, if the average number of nodes infected
by one seed is larger than 1, an epidemic will occur. In discrete time steps, this average number
can be approximately calculated as~\cite{shu2016recovery}
\begin{eqnarray} \label{R}
R&=&
\nonumber
k\sum_{t=1}^{\infty}[(1-\mu)(1-p(0,\alpha))]^{t-1}p(0,\alpha)\\
\nonumber
&+&(k-1)\sum_{t=2}^{\infty}[(1-\mu)(1-p(0,\alpha))]^{t-2}\\
&\times&p(0,\alpha)(1-\mu)[p(1,\alpha)-p(0,\alpha)],
\end{eqnarray}
where the first term of Eq. (\ref{R}) represents the basic reproduction
number without any synergistic effect, the second term denotes the
increment in the basic reproduction number as a result of the synergistic
effect due to the newly infected neighbor, if the seed
indeed successfully infects a neighbor before its recovery. Letting $R=1$ in Eq.~(\ref{R}), we
can approximately calculate the critical invasion threshold as
\begin{eqnarray}\label{qBeta}
\beta_{inv}^{'}=\frac{\mu}{k+(k-1)(1-\mu)\alpha+\mu-1}.
\end{eqnarray}
As shown in Fig.~\ref{fig:ES_regime}(a), the value of $\beta_{inv}^{'}$
agrees well with the value of $\beta^s_{inv}$. For the
case of small initial infected density, the mean-field approximation fails
to capture the dynamical correlation. Due to the synergistic effect, even only one end of the I-I edge transmits the disease
to its susceptible neighbors, the $I_{k,1}$ node becomes $I_{k,2}$, which
has a larger transmission rate than that from the original $I_{k,1}$ node.
As the spreading process continues, more susceptible nodes in the neighborhood
of the infected node are infected so the $I_{k,2}$ nodes become $I_{k,3}$,
$I_{k,3}$ becomes $I_{k,4}$, and so on, leading to a cascading process
that results in explosive spreading.

To gain further insights into the cascading phenomenon and the explosive
increase of $\rho(\infty)$ with $\beta$ for $\alpha>\alpha_c$, we calculate
the fraction $i_m$ of infected nodes with $m$ ($m=0,1,...,k$) infected
neighbors versus time for $\beta$ slightly larger than $\beta_{inv}$ (for
$\alpha=0.9$) and $\beta_c$ (for $\alpha=0$). For $\alpha < \alpha_c$
(e.g., $\alpha=0$), the synergistic SIS spreading is reduced to the classic
SIS dynamics. As shown in the inset of Fig.~\ref{fig:ES_regime}(b), for
$\beta = 0.0114 \agt \beta_c = 0.0112$, $i_m$ increases with $t$ slowly and
tends to a constant for large time. However, for $\alpha=0.9$, if
$\beta=0.0064 \agt \beta_{inv}^{s}=0.0062$, $i_m$ increases fast initially,
reaches a peak at some small value of $m$ (e.g., $m = 0,1$), and then
decreases rapidly. For larger $m$ values (e.g., $m=3,5$),
$i_m$ increases later and faster in reaching the peak. These provide an explanation for the
continuously and relatively slowly increasing behavior of $\rho(\infty)$
for $\alpha<\alpha_c$ and, more importantly, the explosively
increasing behavior of $\rho(\infty)$ with $\beta$ for $\alpha>\alpha_c$.

\begin{figure}
\centering
\includegraphics[width=\linewidth]{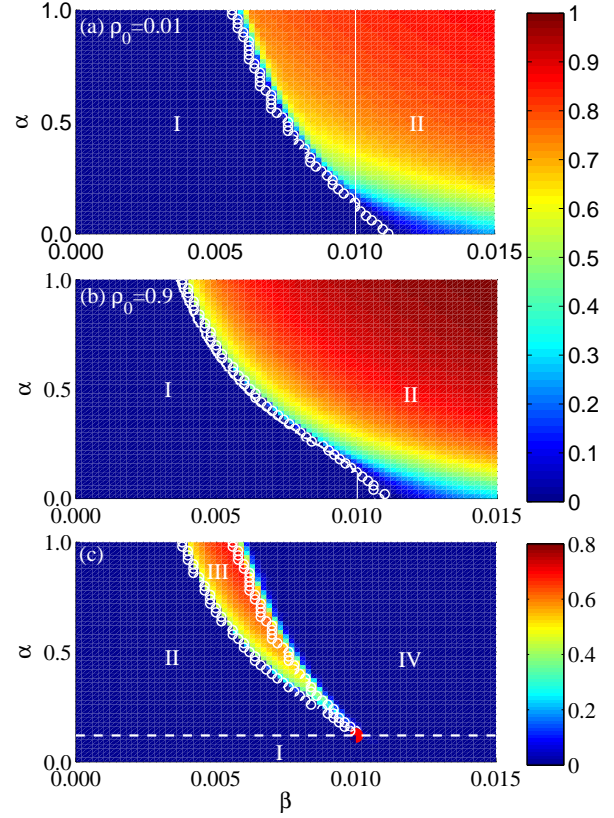}
\caption{(Color online) Steady state infected density $\rho(\infty)$
and region of hysteresis in the parameter plane ($\beta$, $\alpha$).
(a,b) For synergistic SIS spreading dynamics on random regular networks,
color-coded values of $\rho(\infty)$ in the parameter plane
($\beta$,$\alpha$) for $\rho_0=0.01$ and $\rho_0=0.9$, respectively. The
numerically obtained invasion threshold $\beta_{inv}^s$ and persistence
threshold $\beta_{per}^s$ (white circles) in (a) and (b), respectively,
are determined by the susceptible measure $\chi$, and the corresponding
theoretical values (white sold line) are from Eqs.~(\ref{eq:RhoInfety})
and~(\ref{eq:BetaC}). The persistence threshold predicted by the mean-filed
theory matches well with that from simulations, but there is disagreement
for the invasion threshold, as shown in (a,b), where I and II denote the
parameter regions where the disease becomes extinct and an outbreak occurs,
respectively. In (c), the color-coded values represent the
difference between the values of $\rho(\infty)$ in (b) and (a).
There are four regions: in region I there is no hysteresis loop
($\alpha < \alpha_c$), in region III there is a hysteresis behavior, and
regines II and IV specify the borders of the hysteresis loop. Other
parameters are $\mu=0.1$ and $k=10$.}
\label{fig:SSD_HL}
\end{figure}

We further examine the impact of parameters $\beta$ and $\alpha$ on
the synergistic SIS spreading dynamics. Figures~\ref{fig:SSD_HL}(a) and (b) show the value of $\rho(\infty)$ in the ($\beta$, $\alpha$)
plane for $\rho_0=0.01$ and $\rho_0=0.9$, respectively. In (a), the solid
curves represent the analytical predictions of $\beta_{inv}$ versus
$\alpha$ obtained from Eq.~\eqref{eq:BetaC2}, and the circles display the numerical predictions of $\beta_{inv}^{s}$ determined by the susceptible measure, which
increases with $\alpha$. The results in (b) show that the persistence
threshold decreases as $\alpha$ is increased. A heuristic explanation for
these results is that, due to the synergistic effect, there is an increase
in the infection probability $p(m, \alpha)$ between the infected nodes and
their susceptible neighbors, thereby reducing the epidemic threshold
(e.g., $\beta_{inv}$ and $\beta_{per}$). In Figs.~\ref{fig:SSD_HL}(a) and (b), depending on whether the disease becomes extinct or there
is an outbreak, we can divide the parameter plane into regions I and II,
respectively. For $\beta > \beta_{inv}$ (or $\beta > \beta_{per}$),
$\rho(\infty)$ increases with $\alpha$ due to the enhancement in the
transmission rate between the infected node and its susceptible neighbors.
Since the initial fraction of infected seeds impacts only the steady
state associated with the region of the hysteresis loop, we can determine
this region by computing the difference between the values of every
point ($\beta$,$\alpha$) in Figs.~\ref{fig:SSD_HL}(b) and
\ref{fig:SSD_HL}(a). As shown in Fig.~\ref{fig:SSD_HL}(c), there
are four regions. Only when $\alpha$ is larger than a critical value
$\alpha_c$ [obtained from Eqs.~(\ref{eq:RhoInfety}),~(\ref{eq:DiffRho})
and~(\ref{eq:AlphaC})] will the final density $\rho(\infty)$ increase
with $\beta$ explosively (regions II, III, and IV) and a hysteresis loop
appears (region III). Otherwise there is no hysteresis (region I). In region
II, the disease becomes extinct, but there is an outbreak in region IV.

\begin{figure}
\centering
\includegraphics[width=\linewidth]{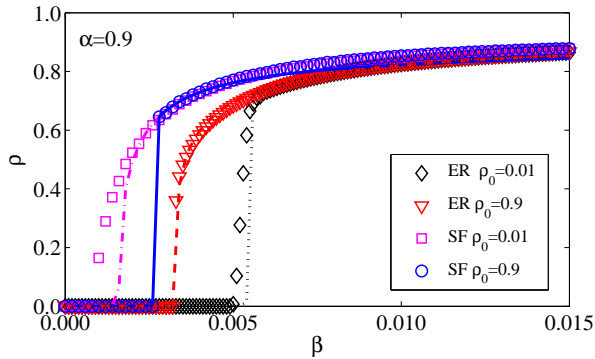}
\caption{(Color online) Synergistic SIS spreading processes on
random and scale-free networks. Steady state density of infected nodes
versus $\beta$: where symbols are results from simulation and the
corresponding lines are predictions of the master equations
Eqs.~(\ref{eq:BetaS})-(\ref{eq:BetaI}). The network parameters are $N=10^4$
and $\langle k \rangle=10$.}
\label{fig:ERSF}
\end{figure}

While we focus our study on RRNs for the reason that
an understanding of explosive spreading can be obtained, the phenomenon
can arise in general complex networks. To demonstrate this, we simulate
synergistic spreading dynamics on Erd\"{o}s-R\'{e}nyi (ER) random and
scale-free networks. Figure~\ref{fig:ERSF} shows, for ER networks, an
explosive increase in the steady state infection density and a hysteresis
loop with the parameter $\beta$. We also investigate the spreading dynamics
on scale-free networks~\cite{Newman:2002} constructed according to the
standard configuration model~\cite{Catanzaro:2005}. The degree distribution is
$P(k)={\Gamma}k^{-\gamma}$, where $\gamma$ is degree exponent and the
coefficient is $\Gamma=1/\sum_{k_{min}}^{k_{max}}k^{-\gamma}$ with the
minimum degree $k_{min}=3$, maximum degree $k_{max}{\sim}N^{1/(\gamma-1)}$ and $\gamma=3.0$.
The phenomena of explosive spreading and hysteresis loop are presented, as
shown in Fig.~\ref{fig:ERSF}.

\section{Discussion} \label{sec:discussion}

Synergy is a ubiquitous phenomenon in biological and social systems, and
one is naturally curious about its effect on spreading dynamics on networks.
There were previous works on synergistic irreversible spreading dynamics,
and the goals of this paper are to construct and analyze a generic model
for synergistic reversible spreading, where the effect of synergy is taken
into account through enhancement in the transmission rate between an infected
node and its susceptible neighbors. There are two factors determining
the synergistic effect: the number of infected neighbors connected to the
infected node that is to transmit the disease to one of its susceptible
neighbors and the strength of the synergistic reinforcement effect.
For RRNs, the synergistic reversible spreading dynamics
can be treated analytically by using the approach of master equations,
as well as a mean field approximation. Qualitatively, we find that synergy
promotes spreading. The manner by which spreading is enhanced is, however,
quite striking. In particular, if the strength is above a critical value that
is solely determined by the degree of the network, there is an explosive
outbreak of the disease in that the steady state infection density increases
abruptly and drastically as the basic transmission rate passes through a
critical value. Associated with the explosive behavior is a hysteresis loop
whereas, if the transmission rate is reduced through a different threshold,
the final infected population collapses to zero. All these results have
been obtained both analytically and numerically. While the analysis is
feasible for RRNs, numerically we find that a similar
explosive behavior occurs for general complex networks with a random or
a scale-free topology.

The main contributions of our work are thus the discovery of synergy
induced explosive outbreak for reversible spreading dynamics, and a
qualitative and quantitative understanding of the phenomenon. A number of questions still remain. For example, the effects of network structural characteristics such as clustering~\cite{Serrano:2006,Newman:2009,
cui2012emergence}, community~\cite{Girvan:2002,Fortunato:2010,
gong2013efficient}, and core-periphery~\cite{borgatti2000models,
holme2005core,liu2014core,verma2016emergence} on synergistic spreading
dynamics need to be studied. The approach of master equations
needs to be improved beyond random regular networks.
Finally, the study needs to be extended to more
realistic networks such as multiplex
networks~\cite{SGR:2014,WTYDLL:2014,Liu:2016,Kivel:2014}, or temporal
networks~\cite{Holme:2012,Barrat:2013,Moinet:2015}.

\acknowledgments
This work was supported by the National Natural Science Foundation of
China under Grants Nos.~11105025, 11575041, and 61433014, and the
Fundamental Research Funds for the Central Universities (Grant
No.~ZYGX2015J153). YCL was supported by ARO under Grant No.~W911NF-14-1-0504.

\bibliographystyle{apsrev4-1}
\bibliography{Synergy_Spreading}

\begin{thebibliography}{63}%
\makeatletter
\providecommand \@ifxundefined [1]{%
 \@ifx{#1\undefined}
}%
\providecommand \@ifnum [1]{%
 \ifnum #1\expandafter \@firstoftwo
 \else \expandafter \@secondoftwo
 \fi
}%
\providecommand \@ifx [1]{%
 \ifx #1\expandafter \@firstoftwo
 \else \expandafter \@secondoftwo
 \fi
}%
\providecommand \natexlab [1]{#1}%
\providecommand \enquote  [1]{``#1''}%
\providecommand \bibnamefont  [1]{#1}%
\providecommand \bibfnamefont [1]{#1}%
\providecommand \citenamefont [1]{#1}%
\providecommand \href@noop [0]{\@secondoftwo}%
\providecommand \href [0]{\begingroup \@sanitize@url \@href}%
\providecommand \@href[1]{\@@startlink{#1}\@@href}%
\providecommand \@@href[1]{\endgroup#1\@@endlink}%
\providecommand \@sanitize@url [0]{\catcode `\\12\catcode `\$12\catcode
  `\&12\catcode `\#12\catcode `\^12\catcode `\_12\catcode `\%12\relax}%
\providecommand \@@startlink[1]{}%
\providecommand \@@endlink[0]{}%
\providecommand \url  [0]{\begingroup\@sanitize@url \@url }%
\providecommand \@url [1]{\endgroup\@href {#1}{\urlprefix }}%
\providecommand \urlprefix  [0]{URL }%
\providecommand \Eprint [0]{\href }%
\providecommand \doibase [0]{http://dx.doi.org/}%
\providecommand \selectlanguage [0]{\@gobble}%
\providecommand \bibinfo  [0]{\@secondoftwo}%
\providecommand \bibfield  [0]{\@secondoftwo}%
\providecommand \translation [1]{[#1]}%
\providecommand \BibitemOpen [0]{}%
\providecommand \bibitemStop [0]{}%
\providecommand \bibitemNoStop [0]{.\EOS\space}%
\providecommand \EOS [0]{\spacefactor3000\relax}%
\providecommand \BibitemShut  [1]{\csname bibitem#1\endcsname}%
\let\auto@bib@innerbib\@empty
\bibitem [{\citenamefont {Barrat}\ \emph {et~al.}(2008)\citenamefont {Barrat},
  \citenamefont {Barthelemy},\ and\ \citenamefont {Vespignani}}]{BBV:book}%
  \BibitemOpen
  \bibfield  {author} {\bibinfo {author} {\bibfnamefont {A.}~\bibnamefont
  {Barrat}}, \bibinfo {author} {\bibfnamefont {M.}~\bibnamefont {Barthelemy}},
  \ and\ \bibinfo {author} {\bibfnamefont {A.}~\bibnamefont {Vespignani}},\
  }\href@noop {} {\emph {\bibinfo {title} {Dynamical processes on complex
  networks}}}\ (\bibinfo  {publisher} {Cambridge University Press},\ \bibinfo
  {address} {Cambridge, UK},\ \bibinfo {year} {2008})\BibitemShut {NoStop}%
\bibitem [{\citenamefont {Castellano}\ \emph {et~al.}(2009)\citenamefont
  {Castellano}, \citenamefont {Fortunato},\ and\ \citenamefont
  {Loreto}}]{Castellano:2009}%
  \BibitemOpen
  \bibfield  {author} {\bibinfo {author} {\bibfnamefont {C.}~\bibnamefont
  {Castellano}}, \bibinfo {author} {\bibfnamefont {S.}~\bibnamefont
  {Fortunato}}, \ and\ \bibinfo {author} {\bibfnamefont {V.}~\bibnamefont
  {Loreto}},\ }\href@noop {} {\bibfield  {journal} {\bibinfo  {journal} {Rev.
  Mod. Phys.}\ }\textbf {\bibinfo {volume} {81}},\ \bibinfo {pages} {591}
  (\bibinfo {year} {2009})}\BibitemShut {NoStop}%
\bibitem [{\citenamefont {Newman}(2010)}]{Newman:book}%
  \BibitemOpen
  \bibfield  {author} {\bibinfo {author} {\bibfnamefont {M.~E.~J.}\
  \bibnamefont {Newman}},\ }\href@noop {} {\emph {\bibinfo {title} {Networks:
  An Introduction}}}\ (\bibinfo  {publisher} {Oxford University Press},\
  \bibinfo {address} {Oxford, UK},\ \bibinfo {year} {2010})\BibitemShut
  {NoStop}%
\bibitem [{\citenamefont {Pastor-Satorras}\ \emph {et~al.}(2015)\citenamefont
  {Pastor-Satorras}, \citenamefont {Castellano}, \citenamefont {Van~Mieghem},\
  and\ \citenamefont {Vespignani}}]{Pastor-Satorras:2015}%
  \BibitemOpen
  \bibfield  {author} {\bibinfo {author} {\bibfnamefont {R.}~\bibnamefont
  {Pastor-Satorras}}, \bibinfo {author} {\bibfnamefont {C.}~\bibnamefont
  {Castellano}}, \bibinfo {author} {\bibfnamefont {P.}~\bibnamefont
  {Van~Mieghem}}, \ and\ \bibinfo {author} {\bibfnamefont {A.}~\bibnamefont
  {Vespignani}},\ }\href@noop {} {\bibfield  {journal} {\bibinfo  {journal}
  {Rev. Mod. Phys.}\ }\textbf {\bibinfo {volume} {87}},\ \bibinfo {pages} {925}
  (\bibinfo {year} {2015})}\BibitemShut {NoStop}%
\bibitem [{\citenamefont {Pastor-Satorras}\ and\ \citenamefont
  {Vespignani}(2001)}]{Pastor-Satorras:2001}%
  \BibitemOpen
  \bibfield  {author} {\bibinfo {author} {\bibfnamefont {R.}~\bibnamefont
  {Pastor-Satorras}}\ and\ \bibinfo {author} {\bibfnamefont {A.}~\bibnamefont
  {Vespignani}},\ }\href@noop {} {\bibfield  {journal} {\bibinfo  {journal}
  {Phys. Rev. Lett.}\ }\textbf {\bibinfo {volume} {86}},\ \bibinfo {pages}
  {3200} (\bibinfo {year} {2001})}\BibitemShut {NoStop}%
\bibitem [{\citenamefont {Newman}(2002)}]{Newman:2002}%
  \BibitemOpen
  \bibfield  {author} {\bibinfo {author} {\bibfnamefont {M.~E.~J.}\
  \bibnamefont {Newman}},\ }\href@noop {} {\bibfield  {journal} {\bibinfo
  {journal} {Phys. Rev. E}\ }\textbf {\bibinfo {volume} {66}},\ \bibinfo
  {pages} {016128} (\bibinfo {year} {2002})}\BibitemShut {NoStop}%
\bibitem [{\citenamefont {Zanette}(2002)}]{Zanette:2002}%
  \BibitemOpen
  \bibfield  {author} {\bibinfo {author} {\bibfnamefont {D.~H.}\ \bibnamefont
  {Zanette}},\ }\href@noop {} {\bibfield  {journal} {\bibinfo  {journal} {Phys.
  Rev. E}\ }\textbf {\bibinfo {volume} {65}},\ \bibinfo {pages} {041908}
  (\bibinfo {year} {2002})}\BibitemShut {NoStop}%
\bibitem [{\citenamefont {Liu}\ \emph {et~al.}(2003)\citenamefont {Liu},
  \citenamefont {Lai},\ and\ \citenamefont {Ye}}]{LLY:2003}%
  \BibitemOpen
  \bibfield  {author} {\bibinfo {author} {\bibfnamefont {Z.}~\bibnamefont
  {Liu}}, \bibinfo {author} {\bibfnamefont {Y.-C.}\ \bibnamefont {Lai}}, \ and\
  \bibinfo {author} {\bibfnamefont {N.}~\bibnamefont {Ye}},\ }\href@noop {}
  {\bibfield  {journal} {\bibinfo  {journal} {Phys. Rev. E}\ }\textbf {\bibinfo
  {volume} {67}},\ \bibinfo {pages} {031911} (\bibinfo {year}
  {2003})}\BibitemShut {NoStop}%
\bibitem [{\citenamefont {Barth\'elemy}\ \emph {et~al.}(2004)\citenamefont
  {Barth\'elemy}, \citenamefont {Barrat}, \citenamefont {Pastor-Satorras},\
  and\ \citenamefont {Vespignani}}]{BBPSV:2004}%
  \BibitemOpen
  \bibfield  {author} {\bibinfo {author} {\bibfnamefont {M.}~\bibnamefont
  {Barth\'elemy}}, \bibinfo {author} {\bibfnamefont {A.}~\bibnamefont
  {Barrat}}, \bibinfo {author} {\bibfnamefont {R.}~\bibnamefont
  {Pastor-Satorras}}, \ and\ \bibinfo {author} {\bibfnamefont {A.}~\bibnamefont
  {Vespignani}},\ }\href {\doibase 10.1103/PhysRevLett.92.178701} {\bibfield
  {journal} {\bibinfo  {journal} {Phys. Rev. Lett.}\ }\textbf {\bibinfo
  {volume} {92}},\ \bibinfo {pages} {178701} (\bibinfo {year}
  {2004})}\BibitemShut {NoStop}%
\bibitem [{\citenamefont {Small}\ and\ \citenamefont {Tse}(2005)}]{Small:2005}%
  \BibitemOpen
  \bibfield  {author} {\bibinfo {author} {\bibfnamefont {M.}~\bibnamefont
  {Small}}\ and\ \bibinfo {author} {\bibfnamefont {C.~K.}\ \bibnamefont
  {Tse}},\ }\href@noop {} {\bibfield  {journal} {\bibinfo  {journal} {Int. J.
  Bif. Chaos}\ }\textbf {\bibinfo {volume} {15}},\ \bibinfo {pages} {1745}
  (\bibinfo {year} {2005})}\BibitemShut {NoStop}%
\bibitem [{\citenamefont {Zhou}\ \emph {et~al.}(2007)\citenamefont {Zhou},
  \citenamefont {Liu},\ and\ \citenamefont {Li}}]{ZLL:2007}%
  \BibitemOpen
  \bibfield  {author} {\bibinfo {author} {\bibfnamefont {J.}~\bibnamefont
  {Zhou}}, \bibinfo {author} {\bibfnamefont {Z.}~\bibnamefont {Liu}}, \ and\
  \bibinfo {author} {\bibfnamefont {B.}~\bibnamefont {Li}},\ }\href@noop {}
  {\bibfield  {journal} {\bibinfo  {journal} {Phys. Lett. A}\ }\textbf
  {\bibinfo {volume} {368}},\ \bibinfo {pages} {458} (\bibinfo {year}
  {2007})}\BibitemShut {NoStop}%
\bibitem [{\citenamefont {Yang}\ \emph {et~al.}(2008)\citenamefont {Yang},
  \citenamefont {Huang},\ and\ \citenamefont {Lai}}]{YHL:2008}%
  \BibitemOpen
  \bibfield  {author} {\bibinfo {author} {\bibfnamefont {R.}~\bibnamefont
  {Yang}}, \bibinfo {author} {\bibfnamefont {L.}~\bibnamefont {Huang}}, \ and\
  \bibinfo {author} {\bibfnamefont {Y.-C.}\ \bibnamefont {Lai}},\ }\href
  {\doibase 10.1103/PhysRevE.78.026111} {\bibfield  {journal} {\bibinfo
  {journal} {Phys. Rev. E}\ }\textbf {\bibinfo {volume} {78}},\ \bibinfo
  {pages} {026111} (\bibinfo {year} {2008})}\BibitemShut {NoStop}%
\bibitem [{\citenamefont {Tang}\ \emph {et~al.}(2009)\citenamefont {Tang},
  \citenamefont {Liu},\ and\ \citenamefont {Li}}]{TLL:2009}%
  \BibitemOpen
  \bibfield  {author} {\bibinfo {author} {\bibfnamefont {M.}~\bibnamefont
  {Tang}}, \bibinfo {author} {\bibfnamefont {Z.}~\bibnamefont {Liu}}, \ and\
  \bibinfo {author} {\bibfnamefont {B.}~\bibnamefont {Li}},\ }\href@noop {}
  {\bibfield  {journal} {\bibinfo  {journal} {Europhys. Lett.}\ }\textbf
  {\bibinfo {volume} {87}},\ \bibinfo {pages} {18005} (\bibinfo {year}
  {2009})}\BibitemShut {NoStop}%
\bibitem [{\citenamefont {Gross}\ \emph {et~al.}(2006)\citenamefont {Gross},
  \citenamefont {D'Lima},\ and\ \citenamefont {Blasius}}]{Gross:2006}%
  \BibitemOpen
  \bibfield  {author} {\bibinfo {author} {\bibfnamefont {T.}~\bibnamefont
  {Gross}}, \bibinfo {author} {\bibfnamefont {C.~J.~D.}\ \bibnamefont
  {D'Lima}}, \ and\ \bibinfo {author} {\bibfnamefont {B.}~\bibnamefont
  {Blasius}},\ }\href {\doibase 10.1103/PhysRevLett.96.208701} {\bibfield
  {journal} {\bibinfo  {journal} {Phys. Rev. Lett.}\ }\textbf {\bibinfo
  {volume} {96}},\ \bibinfo {pages} {208701} (\bibinfo {year}
  {2006})}\BibitemShut {NoStop}%
\bibitem [{\citenamefont {Kitsak}\ \emph {et~al.}(2010)\citenamefont {Kitsak},
  \citenamefont {Gallos}, \citenamefont {Havlin}, \citenamefont {Liljeros},
  \citenamefont {Muchnik}, \citenamefont {Stanley},\ and\ \citenamefont
  {Makse}}]{KGHLMSM:2010}%
  \BibitemOpen
  \bibfield  {author} {\bibinfo {author} {\bibfnamefont {M.}~\bibnamefont
  {Kitsak}}, \bibinfo {author} {\bibfnamefont {L.~K.}\ \bibnamefont {Gallos}},
  \bibinfo {author} {\bibfnamefont {S.}~\bibnamefont {Havlin}}, \bibinfo
  {author} {\bibfnamefont {F.}~\bibnamefont {Liljeros}}, \bibinfo {author}
  {\bibfnamefont {L.}~\bibnamefont {Muchnik}}, \bibinfo {author} {\bibfnamefont
  {H.~E.}\ \bibnamefont {Stanley}}, \ and\ \bibinfo {author} {\bibfnamefont
  {H.~A.}\ \bibnamefont {Makse}},\ }\href@noop {} {\bibfield  {journal}
  {\bibinfo  {journal} {Nat. Phys.}\ }\textbf {\bibinfo {volume} {6}},\
  \bibinfo {pages} {888} (\bibinfo {year} {2010})}\BibitemShut {NoStop}%
\bibitem [{\citenamefont {Yang}\ \emph {et~al.}(2011)\citenamefont {Yang},
  \citenamefont {Wang}, \citenamefont {Lai}, \citenamefont {Xie},\ and\
  \citenamefont {Wang}}]{YWLXW:2011}%
  \BibitemOpen
  \bibfield  {author} {\bibinfo {author} {\bibfnamefont {H.-X.}\ \bibnamefont
  {Yang}}, \bibinfo {author} {\bibfnamefont {W.-X.}\ \bibnamefont {Wang}},
  \bibinfo {author} {\bibfnamefont {Y.-C.}\ \bibnamefont {Lai}}, \bibinfo
  {author} {\bibfnamefont {Y.-B.}\ \bibnamefont {Xie}}, \ and\ \bibinfo
  {author} {\bibfnamefont {B.-H.}\ \bibnamefont {Wang}},\ }\href {\doibase
  10.1103/PhysRevE.84.045101} {\bibfield  {journal} {\bibinfo  {journal} {Phys.
  Rev. E}\ }\textbf {\bibinfo {volume} {84}},\ \bibinfo {pages} {045101}
  (\bibinfo {year} {2011})}\BibitemShut {NoStop}%
\bibitem [{\citenamefont {Zhu}\ \emph {et~al.}(2012)\citenamefont {Zhu},
  \citenamefont {Fu},\ and\ \citenamefont {Chen}}]{ZFC:2012}%
  \BibitemOpen
  \bibfield  {author} {\bibinfo {author} {\bibfnamefont {G.-H.}\ \bibnamefont
  {Zhu}}, \bibinfo {author} {\bibfnamefont {X.-C.}\ \bibnamefont {Fu}}, \ and\
  \bibinfo {author} {\bibfnamefont {G.-R.}\ \bibnamefont {Chen}},\ }\href@noop
  {} {\bibfield  {journal} {\bibinfo  {journal} {Appl. Math. Model.}\ }\textbf
  {\bibinfo {volume} {36}},\ \bibinfo {pages} {5808} (\bibinfo {year}
  {2012})}\BibitemShut {NoStop}%
\bibitem [{\citenamefont {Brockmann}\ and\ \citenamefont
  {Helbing}(2013)}]{BH:2013}%
  \BibitemOpen
  \bibfield  {author} {\bibinfo {author} {\bibfnamefont {D.}~\bibnamefont
  {Brockmann}}\ and\ \bibinfo {author} {\bibfnamefont {D.}~\bibnamefont
  {Helbing}},\ }\href@noop {} {\bibfield  {journal} {\bibinfo  {journal}
  {Science}\ }\textbf {\bibinfo {volume} {342}},\ \bibinfo {pages} {1337}
  (\bibinfo {year} {2013})}\BibitemShut {NoStop}%
\bibitem [{\citenamefont {Gleeson}(2013)}]{Glesson:2013}%
  \BibitemOpen
  \bibfield  {author} {\bibinfo {author} {\bibfnamefont {J.~P.}\ \bibnamefont
  {Gleeson}},\ }\href {\doibase 10.1103/PhysRevX.3.021004} {\bibfield
  {journal} {\bibinfo  {journal} {Phys. Rev. X}\ }\textbf {\bibinfo {volume}
  {3}},\ \bibinfo {pages} {021004} (\bibinfo {year} {2013})}\BibitemShut
  {NoStop}%
\bibitem [{\citenamefont {Boccaletti}\ \emph {et~al.}(2014)\citenamefont
  {Boccaletti}, \citenamefont {Bianconi},\ and\ \citenamefont
  {Criado}}]{SGR:2014}%
  \BibitemOpen
  \bibfield  {author} {\bibinfo {author} {\bibfnamefont {S.}~\bibnamefont
  {Boccaletti}}, \bibinfo {author} {\bibfnamefont {G.}~\bibnamefont
  {Bianconi}}, \ and\ \bibinfo {author} {\bibfnamefont {R.~e.~a.}\ \bibnamefont
  {Criado}},\ }\href@noop {} {\bibfield  {journal} {\bibinfo  {journal} {Phys.
  Rep.}\ }\textbf {\bibinfo {volume} {544}},\ \bibinfo {pages} {1} (\bibinfo
  {year} {2014})}\BibitemShut {NoStop}%
\bibitem [{\citenamefont {Granell}\ \emph {et~al.}(2013)\citenamefont
  {Granell}, \citenamefont {G\'{o}mez},\ and\ \citenamefont
  {Arenas}}]{GGA2013}%
  \BibitemOpen
  \bibfield  {author} {\bibinfo {author} {\bibfnamefont {C.}~\bibnamefont
  {Granell}}, \bibinfo {author} {\bibfnamefont {S.}~\bibnamefont {G\'{o}mez}},
  \ and\ \bibinfo {author} {\bibfnamefont {A.}~\bibnamefont {Arenas}},\
  }\href@noop {} {\bibfield  {journal} {\bibinfo  {journal} {Phys. Rev. Lett.}\
  }\textbf {\bibinfo {volume} {111}},\ \bibinfo {pages} {128701} (\bibinfo
  {year} {2013})}\BibitemShut {NoStop}%
\bibitem [{\citenamefont {Wang}\ \emph {et~al.}(2014)\citenamefont {Wang},
  \citenamefont {Tang}, \citenamefont {Yang}, \citenamefont {Do}, \citenamefont
  {Lai},\ and\ \citenamefont {Lee}}]{WTYDLL:2014}%
  \BibitemOpen
  \bibfield  {author} {\bibinfo {author} {\bibfnamefont {W.}~\bibnamefont
  {Wang}}, \bibinfo {author} {\bibfnamefont {M.}~\bibnamefont {Tang}}, \bibinfo
  {author} {\bibfnamefont {H.}~\bibnamefont {Yang}}, \bibinfo {author}
  {\bibfnamefont {Y.-H.}\ \bibnamefont {Do}}, \bibinfo {author} {\bibfnamefont
  {Y.-C.}\ \bibnamefont {Lai}}, \ and\ \bibinfo {author} {\bibfnamefont
  {G.~W.}\ \bibnamefont {Lee}},\ }\href@noop {} {\bibfield  {journal} {\bibinfo
   {journal} {Sci. Rep.}\ }\textbf {\bibinfo {volume} {4}},\ \bibinfo {pages}
  {5097} (\bibinfo {year} {2014})}\BibitemShut {NoStop}%
\bibitem [{\citenamefont {Liu}\ \emph {et~al.}(2016)\citenamefont {Liu},
  \citenamefont {Wang}, \citenamefont {Tang},\ and\ \citenamefont
  {Zhang}}]{Liu:2016}%
  \BibitemOpen
  \bibfield  {author} {\bibinfo {author} {\bibfnamefont {Q.~H.}\ \bibnamefont
  {Liu}}, \bibinfo {author} {\bibfnamefont {W.}~\bibnamefont {Wang}}, \bibinfo
  {author} {\bibfnamefont {M.}~\bibnamefont {Tang}}, \ and\ \bibinfo {author}
  {\bibfnamefont {H.~F.}\ \bibnamefont {Zhang}},\ }\href@noop {} {\bibfield
  {journal} {\bibinfo  {journal} {Sci. Rep.}\ }\textbf {\bibinfo {volume}
  {6}},\ \bibinfo {pages} {25617} (\bibinfo {year} {2016})}\BibitemShut
  {NoStop}%
\bibitem [{\citenamefont {Bancal}\ and\ \citenamefont
  {Pastor-Satorras}(2010)}]{Bancal:2010}%
  \BibitemOpen
  \bibfield  {author} {\bibinfo {author} {\bibfnamefont {J.-D.}\ \bibnamefont
  {Bancal}}\ and\ \bibinfo {author} {\bibfnamefont {R.}~\bibnamefont
  {Pastor-Satorras}},\ }\href@noop {} {\bibfield  {journal} {\bibinfo
  {journal} {Euro. Phys. J. B}\ }\textbf {\bibinfo {volume} {76}},\ \bibinfo
  {pages} {109} (\bibinfo {year} {2010})}\BibitemShut {NoStop}%
\bibitem [{\citenamefont {Masuda}\ and\ \citenamefont
  {Konno}(2006)}]{Masuda:2006}%
  \BibitemOpen
  \bibfield  {author} {\bibinfo {author} {\bibfnamefont {N.}~\bibnamefont
  {Masuda}}\ and\ \bibinfo {author} {\bibfnamefont {N.}~\bibnamefont {Konno}},\
  }\href@noop {} {\bibfield  {journal} {\bibinfo  {journal} {J. Theo. Biol.}\
  }\textbf {\bibinfo {volume} {243}},\ \bibinfo {pages} {64} (\bibinfo {year}
  {2006})}\BibitemShut {NoStop}%
\bibitem [{\citenamefont {Granovetter}(1978)}]{Granovetter:1978}%
  \BibitemOpen
  \bibfield  {author} {\bibinfo {author} {\bibfnamefont {M.}~\bibnamefont
  {Granovetter}},\ }\href@noop {} {\bibfield  {journal} {\bibinfo  {journal}
  {Ame. J. Soc.}\ }\textbf {\bibinfo {volume} {83}},\ \bibinfo {pages} {1420}
  (\bibinfo {year} {1978})}\BibitemShut {NoStop}%
\bibitem [{\citenamefont {Watts}(2002)}]{Watts:2002}%
  \BibitemOpen
  \bibfield  {author} {\bibinfo {author} {\bibfnamefont {D.~J.}\ \bibnamefont
  {Watts}},\ }\href@noop {} {\bibfield  {journal} {\bibinfo  {journal} {Proc.
  Nat. Acad. Sci. U.S.A.}\ }\textbf {\bibinfo {volume} {99}},\ \bibinfo {pages}
  {5766} (\bibinfo {year} {2002})}\BibitemShut {NoStop}%
\bibitem [{\citenamefont {Lockwood}(1988)}]{Lockwood:1988}%
  \BibitemOpen
  \bibfield  {author} {\bibinfo {author} {\bibfnamefont {J.~L.}\ \bibnamefont
  {Lockwood}},\ }\href@noop {} {\bibfield  {journal} {\bibinfo  {journal} {Ann.
  Rev. Phytopath.}\ }\textbf {\bibinfo {volume} {26}},\ \bibinfo {pages} {93}
  (\bibinfo {year} {1988})}\BibitemShut {NoStop}%
\bibitem [{\citenamefont {Ludlam}\ \emph {et~al.}(2012)\citenamefont {Ludlam},
  \citenamefont {Gibson}, \citenamefont {Otten},\ and\ \citenamefont
  {Gilligan}}]{Jonathan:2011}%
  \BibitemOpen
  \bibfield  {author} {\bibinfo {author} {\bibfnamefont {J.~J.}\ \bibnamefont
  {Ludlam}}, \bibinfo {author} {\bibfnamefont {G.~J.}\ \bibnamefont {Gibson}},
  \bibinfo {author} {\bibfnamefont {W.}~\bibnamefont {Otten}}, \ and\ \bibinfo
  {author} {\bibfnamefont {C.~A.}\ \bibnamefont {Gilligan}},\ }\href@noop {}
  {\bibfield  {journal} {\bibinfo  {journal} {J. Roy. Soc. Interface}\ }\textbf
  {\bibinfo {volume} {9}},\ \bibinfo {pages} {949} (\bibinfo {year}
  {2012})}\BibitemShut {NoStop}%
\bibitem [{\citenamefont {Centola}(2010)}]{Centola:2010}%
  \BibitemOpen
  \bibfield  {author} {\bibinfo {author} {\bibfnamefont {D.}~\bibnamefont
  {Centola}},\ }\href@noop {} {\bibfield  {journal} {\bibinfo  {journal}
  {Science}\ }\textbf {\bibinfo {volume} {329}},\ \bibinfo {pages} {1194}
  (\bibinfo {year} {2010})}\BibitemShut {NoStop}%
\bibitem [{\citenamefont {Wang}\ \emph {et~al.}(2015)\citenamefont {Wang},
  \citenamefont {Tang}, \citenamefont {Zhang},\ and\ \citenamefont
  {Lai}}]{Wang:2015}%
  \BibitemOpen
  \bibfield  {author} {\bibinfo {author} {\bibfnamefont {W.}~\bibnamefont
  {Wang}}, \bibinfo {author} {\bibfnamefont {M.}~\bibnamefont {Tang}}, \bibinfo
  {author} {\bibfnamefont {H.-F.}\ \bibnamefont {Zhang}}, \ and\ \bibinfo
  {author} {\bibfnamefont {Y.-C.}\ \bibnamefont {Lai}},\ }\href@noop {}
  {\bibfield  {journal} {\bibinfo  {journal} {Phys. Rev. E}\ }\textbf {\bibinfo
  {volume} {92}},\ \bibinfo {pages} {012820} (\bibinfo {year}
  {2015})}\BibitemShut {NoStop}%
\bibitem [{\citenamefont {Hodas}\ and\ \citenamefont
  {Lerman}(2014)}]{Hodas:2014}%
  \BibitemOpen
  \bibfield  {author} {\bibinfo {author} {\bibfnamefont {N.~O.}\ \bibnamefont
  {Hodas}}\ and\ \bibinfo {author} {\bibfnamefont {K.}~\bibnamefont {Lerman}},\
  }\href@noop {} {\bibfield  {journal} {\bibinfo  {journal} {Sci. Rep.}\
  }\textbf {\bibinfo {volume} {4}},\ \bibinfo {pages} {4343} (\bibinfo {year}
  {2014})}\BibitemShut {NoStop}%
\bibitem [{\citenamefont {Lu}\ \emph {et~al.}(2011)\citenamefont {Lu},
  \citenamefont {Chen},\ and\ \citenamefont {Zhou}}]{LCZ:2011}%
  \BibitemOpen
  \bibfield  {author} {\bibinfo {author} {\bibfnamefont {L.}~\bibnamefont
  {Lu}}, \bibinfo {author} {\bibfnamefont {D.-B.}\ \bibnamefont {Chen}}, \ and\
  \bibinfo {author} {\bibfnamefont {T.}~\bibnamefont {Zhou}},\ }\href@noop {}
  {\bibfield  {journal} {\bibinfo  {journal} {New J. Phys.}\ }\textbf {\bibinfo
  {volume} {13}},\ \bibinfo {pages} {123005} (\bibinfo {year}
  {2011})}\BibitemShut {NoStop}%
\bibitem [{\citenamefont {Murray}()}]{Murray:2002}%
  \BibitemOpen
  \bibfield  {author} {\bibinfo {author} {\bibfnamefont {J.~D.}\ \bibnamefont
  {Murray}},\ }\href@noop {} {\emph {\bibinfo {title} {Mathematical Biology,
  Vol. 17 of Interdisciplinary Applied Mathematics, 3rd ed.}}}\ (\bibinfo
  {publisher} {Springer, Berlin, 2002})\BibitemShut {NoStop}%
\bibitem [{\citenamefont {Gordon}(2010)}]{Gordon:2010}%
  \BibitemOpen
  \bibfield  {author} {\bibinfo {author} {\bibfnamefont {D.~M.}\ \bibnamefont
  {Gordon}},\ }\href@noop {} {\emph {\bibinfo {title} {Ant encounters:
  interaction networks and colony behavior}}}\ (\bibinfo  {publisher}
  {Princeton University Press},\ \bibinfo {year} {2010})\BibitemShut {NoStop}%
\bibitem [{\citenamefont {Goldenberg}\ \emph {et~al.}(2001)\citenamefont
  {Goldenberg}, \citenamefont {Libai},\ and\ \citenamefont
  {Muller}}]{GLM:2001}%
  \BibitemOpen
  \bibfield  {author} {\bibinfo {author} {\bibfnamefont {J.}~\bibnamefont
  {Goldenberg}}, \bibinfo {author} {\bibfnamefont {B.}~\bibnamefont {Libai}}, \
  and\ \bibinfo {author} {\bibfnamefont {E.}~\bibnamefont {Muller}},\
  }\href@noop {} {\bibfield  {journal} {\bibinfo  {journal} {Mark. Lett.}\
  }\textbf {\bibinfo {volume} {12}},\ \bibinfo {pages} {211} (\bibinfo {year}
  {2001})}\BibitemShut {NoStop}%
\bibitem [{\citenamefont {P{\'e}rez-Reche}\ \emph {et~al.}(2011)\citenamefont
  {P{\'e}rez-Reche}, \citenamefont {Ludlam}, \citenamefont {Taraskin},\ and\
  \citenamefont {Gilligan}}]{Perez-Reche:2011}%
  \BibitemOpen
  \bibfield  {author} {\bibinfo {author} {\bibfnamefont {F.~J.}\ \bibnamefont
  {P{\'e}rez-Reche}}, \bibinfo {author} {\bibfnamefont {J.~J.}\ \bibnamefont
  {Ludlam}}, \bibinfo {author} {\bibfnamefont {S.~N.}\ \bibnamefont
  {Taraskin}}, \ and\ \bibinfo {author} {\bibfnamefont {C.~A.}\ \bibnamefont
  {Gilligan}},\ }\href@noop {} {\bibfield  {journal} {\bibinfo  {journal}
  {Phys. Rev. Lett.}\ }\textbf {\bibinfo {volume} {106}},\ \bibinfo {pages}
  {218701} (\bibinfo {year} {2011})}\BibitemShut {NoStop}%
\bibitem [{\citenamefont {Taraskin}\ and\ \citenamefont
  {P{\'e}rez-Reche}(2013)}]{Taraskin:2013}%
  \BibitemOpen
  \bibfield  {author} {\bibinfo {author} {\bibfnamefont {S.~N.}\ \bibnamefont
  {Taraskin}}\ and\ \bibinfo {author} {\bibfnamefont {F.~J.}\ \bibnamefont
  {P{\'e}rez-Reche}},\ }\href@noop {} {\bibfield  {journal} {\bibinfo
  {journal} {Phys. Rev. E}\ }\textbf {\bibinfo {volume} {88}},\ \bibinfo
  {pages} {062815} (\bibinfo {year} {2013})}\BibitemShut {NoStop}%
\bibitem [{\citenamefont {Broder-Rodgers}\ \emph {et~al.}(2015)\citenamefont
  {Broder-Rodgers}, \citenamefont {P{\'e}rez-Reche},\ and\ \citenamefont
  {Taraskin}}]{Broder-Rodgers:2015}%
  \BibitemOpen
  \bibfield  {author} {\bibinfo {author} {\bibfnamefont {D.}~\bibnamefont
  {Broder-Rodgers}}, \bibinfo {author} {\bibfnamefont {F.~J.}\ \bibnamefont
  {P{\'e}rez-Reche}}, \ and\ \bibinfo {author} {\bibfnamefont {S.~N.}\
  \bibnamefont {Taraskin}},\ }\href@noop {} {\bibfield  {journal} {\bibinfo
  {journal} {Phys. Rev. E}\ }\textbf {\bibinfo {volume} {92}},\ \bibinfo
  {pages} {062814} (\bibinfo {year} {2015})}\BibitemShut {NoStop}%
\bibitem [{\citenamefont {Lindquist}\ \emph {et~al.}(2011)\citenamefont
  {Lindquist}, \citenamefont {Ma}, \citenamefont {Van~den Driessche},\ and\
  \citenamefont {Willeboordse}}]{Lindquist:2011}%
  \BibitemOpen
  \bibfield  {author} {\bibinfo {author} {\bibfnamefont {J.}~\bibnamefont
  {Lindquist}}, \bibinfo {author} {\bibfnamefont {J.}~\bibnamefont {Ma}},
  \bibinfo {author} {\bibfnamefont {P.}~\bibnamefont {Van~den Driessche}}, \
  and\ \bibinfo {author} {\bibfnamefont {F.~H.}\ \bibnamefont {Willeboordse}},\
  }\href@noop {} {\bibfield  {journal} {\bibinfo  {journal} {J. Math. Biol.}\
  }\textbf {\bibinfo {volume} {62}},\ \bibinfo {pages} {143} (\bibinfo {year}
  {2011})}\BibitemShut {NoStop}%
\bibitem [{\citenamefont {Gleeson}(2011)}]{Glesson:2011}%
  \BibitemOpen
  \bibfield  {author} {\bibinfo {author} {\bibfnamefont {J.~P.}\ \bibnamefont
  {Gleeson}},\ }\href@noop {} {\bibfield  {journal} {\bibinfo  {journal} {Phys.
  Rev. Lett.}\ }\textbf {\bibinfo {volume} {107}},\ \bibinfo {pages} {068701}
  (\bibinfo {year} {2011})}\BibitemShut {NoStop}%
\bibitem [{\citenamefont {Yang}\ \emph {et~al.}(2015)\citenamefont {Yang},
  \citenamefont {Tang},\ and\ \citenamefont {Gross}}]{Yang:2015}%
  \BibitemOpen
  \bibfield  {author} {\bibinfo {author} {\bibfnamefont {H.}~\bibnamefont
  {Yang}}, \bibinfo {author} {\bibfnamefont {M.}~\bibnamefont {Tang}}, \ and\
  \bibinfo {author} {\bibfnamefont {T.}~\bibnamefont {Gross}},\ }\href@noop {}
  {\bibfield  {journal} {\bibinfo  {journal} {Sci. Rep.}\ }\textbf {\bibinfo
  {volume} {5}},\ \bibinfo {pages} {13122} (\bibinfo {year}
  {2015})}\BibitemShut {NoStop}%
\bibitem [{\citenamefont {G{\'o}mez-Garde{\~n}es}\ \emph
  {et~al.}(2016)\citenamefont {G{\'o}mez-Garde{\~n}es}, \citenamefont {Lotero},
  \citenamefont {Taraskin},\ and\ \citenamefont
  {P{\'e}rez-Reche}}]{Gomez:2016}%
  \BibitemOpen
  \bibfield  {author} {\bibinfo {author} {\bibfnamefont {J.}~\bibnamefont
  {G{\'o}mez-Garde{\~n}es}}, \bibinfo {author} {\bibfnamefont {L.}~\bibnamefont
  {Lotero}}, \bibinfo {author} {\bibfnamefont {S.}~\bibnamefont {Taraskin}}, \
  and\ \bibinfo {author} {\bibfnamefont {F.}~\bibnamefont {P{\'e}rez-Reche}},\
  }\href@noop {} {\bibfield  {journal} {\bibinfo  {journal} {Sci. Rep.}\
  }\textbf {\bibinfo {volume} {6}} (\bibinfo {year} {2016})}\BibitemShut
  {NoStop}%
\bibitem [{\citenamefont {Ott}(2002)}]{Ott:book}%
  \BibitemOpen
  \bibfield  {author} {\bibinfo {author} {\bibfnamefont {E.}~\bibnamefont
  {Ott}},\ }\href@noop {} {\emph {\bibinfo {title} {Chaos in Dynamical
  Systems}}},\ \bibinfo {edition} {2nd}\ ed.\ (\bibinfo  {publisher} {Cambridge
  University Press},\ \bibinfo {address} {Cambridge, UK},\ \bibinfo {year}
  {2002})\BibitemShut {NoStop}%
\bibitem [{\citenamefont {Strogatz}\ \emph {et~al.}(1994)\citenamefont
  {Strogatz}, \citenamefont {Friedman}, \citenamefont {Mallinckrodt} \emph
  {et~al.}}]{Strogatz:1994}%
  \BibitemOpen
  \bibfield  {author} {\bibinfo {author} {\bibfnamefont {S.}~\bibnamefont
  {Strogatz}}, \bibinfo {author} {\bibfnamefont {M.}~\bibnamefont {Friedman}},
  \bibinfo {author} {\bibfnamefont {A.~J.}\ \bibnamefont {Mallinckrodt}},
  \emph {et~al.},\ }\href@noop {} {\bibfield  {journal} {\bibinfo  {journal}
  {Computer Phys.}\ }\textbf {\bibinfo {volume} {8}},\ \bibinfo {pages} {532}
  (\bibinfo {year} {1994})}\BibitemShut {NoStop}%
\bibitem [{\citenamefont {Ferreira}\ \emph {et~al.}(2012)\citenamefont
  {Ferreira}, \citenamefont {Castellano},\ and\ \citenamefont
  {Pastor-Satorras}}]{Ferreira:2012}%
  \BibitemOpen
  \bibfield  {author} {\bibinfo {author} {\bibfnamefont {S.~C.}\ \bibnamefont
  {Ferreira}}, \bibinfo {author} {\bibfnamefont {C.}~\bibnamefont
  {Castellano}}, \ and\ \bibinfo {author} {\bibfnamefont {R.}~\bibnamefont
  {Pastor-Satorras}},\ }\href@noop {} {\bibfield  {journal} {\bibinfo
  {journal} {Phys. Rev. E}\ }\textbf {\bibinfo {volume} {86}},\ \bibinfo
  {pages} {041125} (\bibinfo {year} {2012})}\BibitemShut {NoStop}%
\bibitem [{\citenamefont {Shu}\ \emph {et~al.}(2015)\citenamefont {Shu},
  \citenamefont {Wang}, \citenamefont {Tang},\ and\ \citenamefont
  {Do}}]{Shu:2015}%
  \BibitemOpen
  \bibfield  {author} {\bibinfo {author} {\bibfnamefont {P.}~\bibnamefont
  {Shu}}, \bibinfo {author} {\bibfnamefont {W.}~\bibnamefont {Wang}}, \bibinfo
  {author} {\bibfnamefont {M.}~\bibnamefont {Tang}}, \ and\ \bibinfo {author}
  {\bibfnamefont {Y.}~\bibnamefont {Do}},\ }\href@noop {} {\bibfield  {journal}
  {\bibinfo  {journal} {Chaos}\ }\textbf {\bibinfo {volume} {25}},\ \bibinfo
  {pages} {063104} (\bibinfo {year} {2015})}\BibitemShut {NoStop}%
\bibitem [{\citenamefont {Shu}\ \emph {et~al.}(2016)\citenamefont {Shu},
  \citenamefont {Wang}, \citenamefont {Tang}, \citenamefont {Zhao},\ and\
  \citenamefont {Zhang}}]{shu2016recovery}%
  \BibitemOpen
  \bibfield  {author} {\bibinfo {author} {\bibfnamefont {P.}~\bibnamefont
  {Shu}}, \bibinfo {author} {\bibfnamefont {W.}~\bibnamefont {Wang}}, \bibinfo
  {author} {\bibfnamefont {M.}~\bibnamefont {Tang}}, \bibinfo {author}
  {\bibfnamefont {P.}~\bibnamefont {Zhao}}, \ and\ \bibinfo {author}
  {\bibfnamefont {Y.-C.}\ \bibnamefont {Zhang}},\ }\href@noop {} {\bibfield
  {journal} {\bibinfo  {journal} {Chaos}\ }\textbf {\bibinfo {volume} {26}},\
  \bibinfo {pages} {063108} (\bibinfo {year} {2016})}\BibitemShut {NoStop}%
\bibitem [{\citenamefont {Catanzaro}\ \emph {et~al.}(2005)\citenamefont
  {Catanzaro}, \citenamefont {Bogu{\~n}{\'a}},\ and\ \citenamefont
  {Pastor-Satorras}}]{Catanzaro:2005}%
  \BibitemOpen
  \bibfield  {author} {\bibinfo {author} {\bibfnamefont {M.}~\bibnamefont
  {Catanzaro}}, \bibinfo {author} {\bibfnamefont {M.}~\bibnamefont
  {Bogu{\~n}{\'a}}}, \ and\ \bibinfo {author} {\bibfnamefont {R.}~\bibnamefont
  {Pastor-Satorras}},\ }\href@noop {} {\bibfield  {journal} {\bibinfo
  {journal} {Phys. Rev. E}\ }\textbf {\bibinfo {volume} {71}},\ \bibinfo
  {pages} {027103} (\bibinfo {year} {2005})}\BibitemShut {NoStop}%
\bibitem [{\citenamefont {Serrano}\ and\ \citenamefont
  {Bogu\~n\'a}(2006)}]{Serrano:2006}%
  \BibitemOpen
  \bibfield  {author} {\bibinfo {author} {\bibfnamefont {M.~A.}\ \bibnamefont
  {Serrano}}\ and\ \bibinfo {author} {\bibfnamefont {M.}~\bibnamefont
  {Bogu\~n\'a}},\ }\href {\doibase 10.1103/PhysRevLett.97.088701} {\bibfield
  {journal} {\bibinfo  {journal} {Phys. Rev. Lett.}\ }\textbf {\bibinfo
  {volume} {97}},\ \bibinfo {pages} {088701} (\bibinfo {year}
  {2006})}\BibitemShut {NoStop}%
\bibitem [{\citenamefont {Newman}(2009)}]{Newman:2009}%
  \BibitemOpen
  \bibfield  {author} {\bibinfo {author} {\bibfnamefont {M.~E.~J.}\
  \bibnamefont {Newman}},\ }\href@noop {} {\bibfield  {journal} {\bibinfo
  {journal} {Phys. Rev. Lett.}\ }\textbf {\bibinfo {volume} {103}},\ \bibinfo
  {pages} {058701} (\bibinfo {year} {2009})}\BibitemShut {NoStop}%
\bibitem [{\citenamefont {Cui}\ \emph {et~al.}(2012)\citenamefont {Cui},
  \citenamefont {Zhang}, \citenamefont {Tang}, \citenamefont {Hui},\ and\
  \citenamefont {Fu}}]{cui2012emergence}%
  \BibitemOpen
  \bibfield  {author} {\bibinfo {author} {\bibfnamefont {A.-X.}\ \bibnamefont
  {Cui}}, \bibinfo {author} {\bibfnamefont {Z.-K.}\ \bibnamefont {Zhang}},
  \bibinfo {author} {\bibfnamefont {M.}~\bibnamefont {Tang}}, \bibinfo {author}
  {\bibfnamefont {P.~M.}\ \bibnamefont {Hui}}, \ and\ \bibinfo {author}
  {\bibfnamefont {Y.}~\bibnamefont {Fu}},\ }\href@noop {} {\bibfield  {journal}
  {\bibinfo  {journal} {PloS ONE}\ }\textbf {\bibinfo {volume} {7}},\ \bibinfo
  {pages} {e50702} (\bibinfo {year} {2012})}\BibitemShut {NoStop}%
\bibitem [{\citenamefont {Girvan}\ and\ \citenamefont
  {Newman}(2002)}]{Girvan:2002}%
  \BibitemOpen
  \bibfield  {author} {\bibinfo {author} {\bibfnamefont {M.}~\bibnamefont
  {Girvan}}\ and\ \bibinfo {author} {\bibfnamefont {M.~E.}\ \bibnamefont
  {Newman}},\ }\href@noop {} {\bibfield  {journal} {\bibinfo  {journal} {Proc.
  Nat. Acad. Sci. U.S.A.}\ }\textbf {\bibinfo {volume} {99}},\ \bibinfo {pages}
  {7821} (\bibinfo {year} {2002})}\BibitemShut {NoStop}%
\bibitem [{\citenamefont {Fortunato}(2010)}]{Fortunato:2010}%
  \BibitemOpen
  \bibfield  {author} {\bibinfo {author} {\bibfnamefont {S.}~\bibnamefont
  {Fortunato}},\ }\href@noop {} {\bibfield  {journal} {\bibinfo  {journal}
  {Phys. Rep.}\ }\textbf {\bibinfo {volume} {486}},\ \bibinfo {pages} {75}
  (\bibinfo {year} {2010})}\BibitemShut {NoStop}%
\bibitem [{\citenamefont {Gong}\ \emph {et~al.}(2013)\citenamefont {Gong},
  \citenamefont {Tang}, \citenamefont {Hui}, \citenamefont {Zhang},
  \citenamefont {Do},\ and\ \citenamefont {Lai}}]{gong2013efficient}%
  \BibitemOpen
  \bibfield  {author} {\bibinfo {author} {\bibfnamefont {K.}~\bibnamefont
  {Gong}}, \bibinfo {author} {\bibfnamefont {M.}~\bibnamefont {Tang}}, \bibinfo
  {author} {\bibfnamefont {P.~M.}\ \bibnamefont {Hui}}, \bibinfo {author}
  {\bibfnamefont {H.~F.}\ \bibnamefont {Zhang}}, \bibinfo {author}
  {\bibfnamefont {Y.}~\bibnamefont {Do}}, \ and\ \bibinfo {author}
  {\bibfnamefont {Y.-C.}\ \bibnamefont {Lai}},\ }\href@noop {} {\bibfield
  {journal} {\bibinfo  {journal} {PloS ONE}\ }\textbf {\bibinfo {volume} {8}},\
  \bibinfo {pages} {e83489} (\bibinfo {year} {2013})}\BibitemShut {NoStop}%
\bibitem [{\citenamefont {Borgatti}\ and\ \citenamefont
  {Everett}(2000)}]{borgatti2000models}%
  \BibitemOpen
  \bibfield  {author} {\bibinfo {author} {\bibfnamefont {S.~P.}\ \bibnamefont
  {Borgatti}}\ and\ \bibinfo {author} {\bibfnamefont {M.~G.}\ \bibnamefont
  {Everett}},\ }\href@noop {} {\bibfield  {journal} {\bibinfo  {journal} {Soc.
  Net.}\ }\textbf {\bibinfo {volume} {21}},\ \bibinfo {pages} {375} (\bibinfo
  {year} {2000})}\BibitemShut {NoStop}%
\bibitem [{\citenamefont {Holme}(2005)}]{holme2005core}%
  \BibitemOpen
  \bibfield  {author} {\bibinfo {author} {\bibfnamefont {P.}~\bibnamefont
  {Holme}},\ }\href@noop {} {\bibfield  {journal} {\bibinfo  {journal} {Phys.
  Rev. E}\ }\textbf {\bibinfo {volume} {72}},\ \bibinfo {pages} {046111}
  (\bibinfo {year} {2005})}\BibitemShut {NoStop}%
\bibitem [{\citenamefont {Liu}\ \emph {et~al.}(2015)\citenamefont {Liu},
  \citenamefont {Tang}, \citenamefont {Zhou},\ and\ \citenamefont
  {Do}}]{liu2014core}%
  \BibitemOpen
  \bibfield  {author} {\bibinfo {author} {\bibfnamefont {Y.}~\bibnamefont
  {Liu}}, \bibinfo {author} {\bibfnamefont {M.}~\bibnamefont {Tang}}, \bibinfo
  {author} {\bibfnamefont {T.}~\bibnamefont {Zhou}}, \ and\ \bibinfo {author}
  {\bibfnamefont {Y.}~\bibnamefont {Do}},\ }\href@noop {} {\bibfield  {journal}
  {\bibinfo  {journal} {Sci. Rep.}\ }\textbf {\bibinfo {volume} {5}},\ \bibinfo
  {pages} {9602} (\bibinfo {year} {2015})}\BibitemShut {NoStop}%
\bibitem [{\citenamefont {Verma}\ \emph {et~al.}(2016)\citenamefont {Verma},
  \citenamefont {Russmann}, \citenamefont {Ara{\'u}jo}, \citenamefont
  {Nagler},\ and\ \citenamefont {Herrmann}}]{verma2016emergence}%
  \BibitemOpen
  \bibfield  {author} {\bibinfo {author} {\bibfnamefont {T.}~\bibnamefont
  {Verma}}, \bibinfo {author} {\bibfnamefont {F.}~\bibnamefont {Russmann}},
  \bibinfo {author} {\bibfnamefont {N.}~\bibnamefont {Ara{\'u}jo}}, \bibinfo
  {author} {\bibfnamefont {J.}~\bibnamefont {Nagler}}, \ and\ \bibinfo {author}
  {\bibfnamefont {H.}~\bibnamefont {Herrmann}},\ }\href@noop {} {\bibfield
  {journal} {\bibinfo  {journal} {Nat. Commun.}\ }\textbf {\bibinfo {volume}
  {7}} (\bibinfo {year} {2016})}\BibitemShut {NoStop}%
\bibitem [{\citenamefont {Kivel{\"a}}\ \emph {et~al.}(2014)\citenamefont
  {Kivel{\"a}}, \citenamefont {Arenas}, \citenamefont {Barthelemy},
  \citenamefont {Gleeson}, \citenamefont {Moreno},\ and\ \citenamefont
  {Porter}}]{Kivel:2014}%
  \BibitemOpen
  \bibfield  {author} {\bibinfo {author} {\bibfnamefont {M.}~\bibnamefont
  {Kivel{\"a}}}, \bibinfo {author} {\bibfnamefont {A.}~\bibnamefont {Arenas}},
  \bibinfo {author} {\bibfnamefont {M.}~\bibnamefont {Barthelemy}}, \bibinfo
  {author} {\bibfnamefont {J.~P.}\ \bibnamefont {Gleeson}}, \bibinfo {author}
  {\bibfnamefont {Y.}~\bibnamefont {Moreno}}, \ and\ \bibinfo {author}
  {\bibfnamefont {M.~A.}\ \bibnamefont {Porter}},\ }\href@noop {} {\bibfield
  {journal} {\bibinfo  {journal} {J. Comp. Net.}\ }\textbf {\bibinfo {volume}
  {2}},\ \bibinfo {pages} {203} (\bibinfo {year} {2014})}\BibitemShut {NoStop}%
\bibitem [{\citenamefont {Holme}\ and\ \citenamefont
  {Saram{\"a}ki}(2012)}]{Holme:2012}%
  \BibitemOpen
  \bibfield  {author} {\bibinfo {author} {\bibfnamefont {P.}~\bibnamefont
  {Holme}}\ and\ \bibinfo {author} {\bibfnamefont {J.}~\bibnamefont
  {Saram{\"a}ki}},\ }\href@noop {} {\bibfield  {journal} {\bibinfo  {journal}
  {Phys. Rep.}\ }\textbf {\bibinfo {volume} {519}},\ \bibinfo {pages} {97}
  (\bibinfo {year} {2012})}\BibitemShut {NoStop}%
\bibitem [{\citenamefont {Barrat}\ \emph {et~al.}(2013)\citenamefont {Barrat},
  \citenamefont {Fernandez}, \citenamefont {Lin},\ and\ \citenamefont
  {Young}}]{Barrat:2013}%
  \BibitemOpen
  \bibfield  {author} {\bibinfo {author} {\bibfnamefont {A.}~\bibnamefont
  {Barrat}}, \bibinfo {author} {\bibfnamefont {B.}~\bibnamefont {Fernandez}},
  \bibinfo {author} {\bibfnamefont {K.~K.}\ \bibnamefont {Lin}}, \ and\
  \bibinfo {author} {\bibfnamefont {L.-S.}\ \bibnamefont {Young}},\ }\href@noop
  {} {\bibfield  {journal} {\bibinfo  {journal} {Phys. Rev. Lett.}\ }\textbf
  {\bibinfo {volume} {110}},\ \bibinfo {pages} {158702} (\bibinfo {year}
  {2013})}\BibitemShut {NoStop}%
\bibitem [{\citenamefont {Moinet}\ \emph {et~al.}(2015)\citenamefont {Moinet},
  \citenamefont {Starnini},\ and\ \citenamefont
  {Pastor-Satorras}}]{Moinet:2015}%
  \BibitemOpen
  \bibfield  {author} {\bibinfo {author} {\bibfnamefont {A.}~\bibnamefont
  {Moinet}}, \bibinfo {author} {\bibfnamefont {M.}~\bibnamefont {Starnini}}, \
  and\ \bibinfo {author} {\bibfnamefont {R.}~\bibnamefont {Pastor-Satorras}},\
  }\href@noop {} {\bibfield  {journal} {\bibinfo  {journal} {Phys. Rev. Lett.}\
  }\textbf {\bibinfo {volume} {114}},\ \bibinfo {pages} {108701} (\bibinfo
  {year} {2015})}\BibitemShut {NoStop}%
\end{thebibliography}%

\end{document}